\documentclass[12pt]{article}
\usepackage{setspace}
\usepackage{color}
\usepackage[pctex32]{graphics}
\usepackage{graphicx}
\newcommand{\ra}{\rangle }
\newcommand{\la}{\langle }
\newcommand{\ket}[1]{| #1 \rangle }
\newcommand{\bra}[1]{\langle #1 | }

\newcommand{\amp }[2]{\langle #1|#2 \rangle }
\newcommand{\weakv}[3]{ \frac{\langle #1|#2| #3\rangle}{\langle #1 | #3 \rangle }}
\newcommand{\beq}{\begin{equation}}
\newcommand{\eeq}{\end{equation}}

\newcommand{\upa}{\uparrow}
\newcommand{\dwa}{\downarrow}
\newcommand{\beqa}{\begin{eqnarray}}
\newcommand{\eeqa}{\end{eqnarray}}
\newcommand{\beqar}{\begin{eqnarray*}}
\newcommand{\eeqar}{\end{eqnarray*}}

\def \la {\langle}
\def \ra {\rangle}

\def \d {\delta}

\def \u {\uparrow}
\def \d {\downarrow}

\newcommand{\ltwid}{\mathrel{\raise.3ex\hbox{$<$\kern-.75em\lower1ex\hbox{$\sim$}}}}
\newcommand{\gtwid}{\mathrel{\raise.3ex\hbox{$>$\kern-.75em\lower1ex\hbox{$\sim$}}}}
\input colornam.sty
\input epsf
\begin{document}

\centerline{ \bf\Large Pre- and post-selection, weak values, and contextuality}

\bigskip

\centerline{\small   Jeff Tollaksen, Department of Physics and }
\centerline{\small Department of Computational and Data Sciences, }
\centerline{College of Science, George Mason University, Fairfax, VA 22030\footnote{email: jtollaks@gmu.edu}}

\begin{abstract} 
By analyzing 
the concept of contextuality (Bell-Kochen-Specker) in terms of pre-and-post-selection (PPS), 
it is possible to assign  definite values to observables in a new and surprising way.  Physical reasons are presented for restrictions on these assignments.  When measurements are performed which do not disturb the pre- and post-selection (i.e. weak measurements), then novel {\it experimental} aspects of contextuality can be demonstrated including a proof that every PPS-paradox with definite predictions implies contextuality. Certain results of these measurements (eccentric weak values with e.g. negative values outside the spectrum), however,  cannot be explained by a ``classical-like" hidden variable theory.

\end{abstract}

\tableofcontents
\section{\textcolor{black}{\bf  INTRODUCTION}}

A traditional concept of the quantum state $\ket{\Psi_{\mathrm{in}}}$ is that it generally provides only statistical information about the outcome of an Ideal Measurement (IM).  Therefore many authors have proposed that the quantum state could be ``completed" by a hidden-variable-theory (HVT).
A natural category of HVT assigns definite values to all possible observables of a system at all times  has a similar relationship to Quantum Mechanics (QM) as Classical Mechanics (CM) has to Classical Statistical Mechanics (CSM).  I.e. QM can be understood in terms of a deeper theory, the HVT.  The relationship between CM and CSM  is relatively simple because an ideal classical measurement precisely measures a property of a system,  {\it without} affecting the system under study.  Measurement of one property will not interfere with measurement of another property (i.e. measurement does not depend on context) and  
the state of the system can then be said to exist when we are not observing it. There is thus a simple relation between the theory and underlying physical processes (the outcome of measurements directly tells us what values to assign to all variables of the theory). 

Moving over to QM, 
there are two general constraints on any HVT which reproduces QM:  a) the Bell-Kochen-Specker theorem (BKS)~\cite{Bell2,kochen}  and Gleason's~\cite{gleason} theorem  showed that any HVT must be contextual and b) 
Bell's theorem~\cite{Bell1}, showed that any HVT must be non-local.  
Gleason and BKS proved that one cannot assign unique answers to yes-no questions (posed to single systems) in such a way that one can think that  measurement  simply reveals the answer as a pre-existing property  that was intrinsic solely to the quantum system itself.
Some versions of BKS depend only on the structure of observables, while some versions also rely on the state and we therefore define a ``value function"
$V_{\vec{\psi}}(\hat{A})$ (the specification
of the HVT) as the assignment of a value to an observable $\hat{A}$ when
an individual system is in the state $\vec{\psi}$.  
BKS assumed that $V_{\vec{\psi}}(\hat{A})$ should satisfy:
\beq
V_{\vec{\psi}}(F\{\hat{A}\})=F\{V_{\vec{\psi}}(\hat{A})\}
\label{KS1}
\eeq
That is, any functional relation of an operator that is a member of a commuting subset of
observables must also be satisfied if one substitutes the values for the observables into the
functional relations. E.g. if a system is characterized by commuting observables $\hat{A}_1$ and  $\hat{A}_2$  then condition \ref{KS1}  requires 
 that all the {\it relationships or functions} between these operators should also be  satisfied when $V_{\vec{\psi}}(\hat{A}_1)$ and $V_{\vec{\psi}}(\hat{A}_2)$  are substituted into the same functional relations.
This condition determines the sum and product rules:
\begin{eqnarray}
& \,&\, V_{\vec{\psi}}(\hat{A}_1\,+\hat{A}_2)
=V_{\vec{\psi}}(\hat{A}_1)+V_{\vec{\psi}}(\hat{A}_2) \label{sumrule}\\
&\, & \, V_{\vec{\psi}}(\hat{A}_1 \hat{A_2})=V_{\vec{\psi}}(\hat{A}_1)V_{\vec{\psi}}(\hat{A}_2)
\label{prodrule}
\end{eqnarray}
HVTs which  meet these conditions  are {\bf noncontextual}: all yes-no questions can be associated with a value assignment $V_{\vec{\psi}}$ which provides a single unique answer, irrespective of the set of other commuting yes-no questions that it is associated with.
BKS showed that attempting such an assignment to some observables is inconsistent:
in any system (of dimension greater than 2) the $2^n$ possible ``yes-no" assignments (to the $n$ projection operators representing the yes-no questions) cannot be compatible with the sum \ref{sumrule} and product \ref{prodrule} rules  for all orthogonal resolutions of the identity.  
E.g., consider a complete set of spectral projectors of an operator $\hat{A}$ with discrete eigenvalues, so that $\hat{P}_i=\hat{P}_{A=a_i}$, such that 
$\sum_{i=1}^{n}V_{\vec{\psi}}(\hat{P_i})=1$.  
Then only one of the projection operators can give a ``yes" assignment ($\hat{P_i}=1$) and the rest have to be ``no" (i.e.  $\hat{P_i}=0)$. 
However, $\vec{\psi}$ can be decomposed into many different basis sets, and the value that 
$V_{\vec{\psi}}$ assigns must be independent of the particular basis. 
BKS showed that this cannot be done.

The principal result of this article is to question whether BKS is just a formal result (i.e. negative statements concerning the impossibility of a classical-like ``non-contextual-HVTs") or if BKS has new positive aspects including experimental consequences.   
We probe this 
by utilizing the natural connection between counterfactual statements and pre-and-post-selection (PPS). We then connect contextuality with issues that can be probed experimentally by weak measurements.  
PPSs were originally probed with the time-symmetric re-formulation of Quantum Mechanics (TSQM, introduced by 
Aharonov, Bergmann and Lebowitz a/k/a ABL~\cite{abl}).   
 To be useful and interesting,  any re-formulation of QM should meet several criteria such as those met by TSQM: 
\begin{itemize}
\item  TSQM is consistent with all the predictions made by standard QM,  
\item  TSQM brings out features in QM that were missed before: e.g., ABL  considered measurement situations {\em between} two successive IMs in which the transition from a  state $\ket{\Psi_{\mathrm{in}}}$ (pre-selected at a time $t_{\mathrm{in}}$)  to a  state $\ket{\Psi_{\mathrm{fin}}}$ (post-selected at a later time $t_{\mathrm{fin}}$) is  generally disturbed by an intermediate precise measurement.   Post-selection reflects a unique aspect of QM in that measurement results are not determined by equations of motion and initial conditions.  
A subsequent theoretical development arising out of the ABL work was the introduction of the ``Weak Value" (WV) of an observable which was probed by a  new type of quantum measurement called the ``Weak Measurement" (WM) \cite{av}. WM experiments have been performed and results are in very good agreement with theoretical predictions,
\item  TSQM lead to simplifications in calculations (as occurred with the Feynman re-formulation) and stimulated discoveries in other fields: e.g. ABL influenced work in cosmology (e.g. Gell-Mann  and Hartle~\cite{gellman}); in superluminal tunneling (Chiao~\cite{chiao} and Steinberg~\cite{Stein}); in quantum information (e.g. the quantum random walk~\cite{adz} or cryptography~\cite{bub,br}), etc.
\item TSQM suggests generalizations of QM that were missed before - e.g. a new solution~\cite{gross} to the quantum measurement problem.
\end{itemize}

\noindent Using TSQM,  
we show how to assign  definite values to sets of BKS observables in a new and surprising way.  We also show how 
measurement disturbance can arise in a new way when value assignments depend  on both the pre- and post-selection.
An ``intriguing" physical reason is presented to explain why this scheme cannot be applied to 2 or more IMs: the  2 IMs interfere with each other because some assignment of eigenvalues to operators are based on just one of the two vectors (i.e. either the pre- {\it or} the post-selected vector) while some assignment of eigenvalues are based on both vectors (i.e. both the pre-selected {\it and} the post-selected vectors, what we call diagonal-PPS). 
In addition, we show that when measurements are performed which do not disturb the PPS (WMs), then novel {\it experimental} aspects of contextuality can be demonstrated, including a proof that every PPS-paradox with definite predictions implies contextuality. 
We also demonstrate an isomorphism between WVs in BKS situations and EPR entanglement.
Certain results of WMs (eccentric WVs with outside the eigenvalue spectrum), however,  cannot be explained by a ``classical-like" HVT.

\section{\bf PPS, CONTEXTUALITY, AND THE 3-BOX-PARADOX}
\label{3boxesintro}
 
The 3-box-PPS-paradox (\cite{AAD}, verified experimentally ~\cite{stein1}) uses a {\it single} quantum
particle
that is placed in a superposition of being in 3 closed, separated boxes. 
The particle is  pre-selected in $|\Psi_{\mathrm{in}} \rangle 
= 1/\sqrt 3 ~(|A\rangle + |B\rangle +|C\rangle)$  , where $|A\rangle$, $|B\rangle$ and $|C\rangle$
denote the particle localized in boxes $A$, $B$, or $C$,  
respectively.  The particle 
is post-selected in the state  $|\Psi_{\mathrm{fin}} \rangle = 1/\sqrt 3   ~(|A\rangle + 
|B\rangle -|C\rangle)$.
If an ideal (i.e. von Neumann) measurement is performed on box $A$ in the intermediate time (e.g. we open the box), then the 
probability to find the particle in box $A$ is 1, i.e. $\mathrm{\bf \hat{P}}_{\mathrm{A}}=|A\rangle\la A|=1$, given by ABL~\cite{abl}\footnote{The time-symmetry in ABL can already be seen from 
the Born formula
$Prob(a_j,t |\Psi_{\mathrm{in}},t_{\mathrm{in}}) = | \bra{a_j} U(t,t_{\mathrm{in}}) \ket{\Psi_{\mathrm{in}}}|^2$.
Instead of $\textcolor{RedViolet}{\mid\langle a_{1}\!\mid exp({\bf -}iH\Delta
t)\Psi_{\mathrm{in}}\rangle\mid^{2}}$ for a particular outcome $a_{1}$, one may equivalently say that the probability is 
$\textcolor{BlueViolet}{\mid\langle a_{1} exp({\bf +}iH\Delta
t)\mid\!\Psi_{\mathrm{in}}\rangle\mid^{2}}$, i.e. that one applies the time evolution operator to evolve
$a_{1}$ from $t$ to $t_{\mathrm{in}}$, 
 which is the time reverse of the first
picture.  }:
\beq
Prob(a_j,t|\Psi_{\mathrm{in}},t_{\mathrm{in}}; \Psi_{\mathrm{fin}},t_{\mathrm{fin}})  =  \frac{ |\amp{\Psi_{\mathrm{fin}} ,t}{a_j}\amp{a_j}{\Psi_{\mathrm{in}},t}|^2  }{\sum_{a'} |\amp{\Psi_{\mathrm{fin}} ,t}{a'}\amp{a'}{\Psi_{\mathrm{in}},t}|^2}
\label{ablf}
\eeq
which in this case yields, $Prob(\mathrm{\bf \hat{P}}_{A})=
{\vert\langle\Psi_{\mathrm{fin}}\vert \mathrm{\bf \hat{P}}_{A}\vert\Psi_{\mathrm{in}}\rangle\vert^{2}
\over
\vert\langle\Psi_{\mathrm{fin}}\vert
  \mathrm{\bf \hat{P}}_{A}\vert\Psi_{\mathrm{in}}\rangle\vert^{2}+\vert\langle\Psi_{\mathrm{fin}}\vert
\mathrm{\bf \hat{P}}_{B}+\mathrm{\bf \hat{P}}_{C}\vert\Psi_{\mathrm{in}}\rangle\vert^{2}}
=1$.  
This can also be seen intuitively without using ABL: if the particle is not found 
in box $A$, then the initial state $|\Psi_{\mathrm{in}} \rangle$ would be projected 
onto $ 1/\sqrt 2 ~( |B\rangle +
|C\rangle)$, but this is orthogonal to the final state $|\Psi_{\mathrm{fin}}\rangle$.  Therefore the particle
must be found in box $A$.  
Similarly, the probability to 
find the particle in box $B$ is  $1$, i.e. $Prob(\mathrm{\bf \hat{P}}_{\mathrm{B}}=1)=1$, and this is the essence of the ``paradox".  

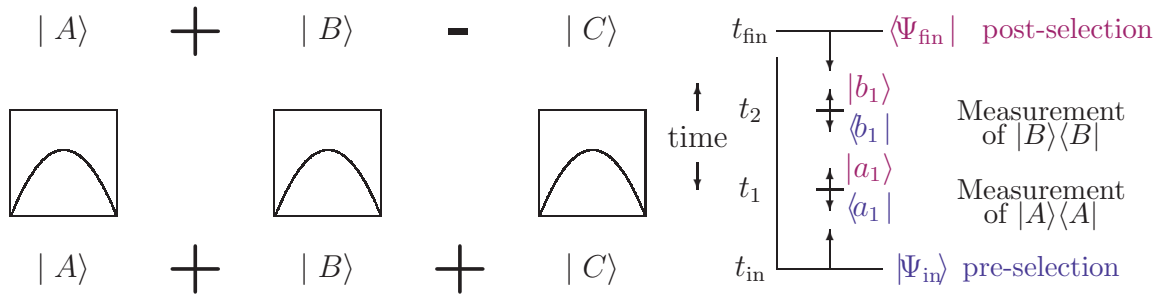
\begin{figure}[h]
\begin{picture}(200,150)(0,0)
\put(0,20){\framebox(40,40)}
\put(100,20){\framebox(40,40)}
\put(200,20){\framebox(40,40)}
\put(20,0){\makebox(0,0){$\mid A \rangle$}}
\put(70,0){\makebox(0,0){\huge\bf+}}
\put(120,0){\makebox(0,0){$\mid B \rangle$}}
\put(170,0){\makebox(0,0){\huge\bf+}}
\put(220,0){\makebox(0,0){$\mid C \rangle$}}

\put(20,90){\makebox(0,0){$\mid A \rangle$}}
\put(70,90){\makebox(0,0){\huge\bf+}}
\put(120,90){\makebox(0,0){$\mid B \rangle$}}
\put(170,90){\makebox(0,0){\huge\bf-}}
\put(220,90){\makebox(0,0){$\mid C \rangle$}}

\bezier{500}(0,20)(20,70)(40,20)
\bezier{500}(100,20)(120,70)(140,20)
\bezier{500}(200,20)(220,70)(240,20)

\put(290,0){\line(0,1){80}}
\put(290,0){\line(1,0){40}}
\put(305,60){\line(1,0){10}}
\put(310,60){\vector(0,1){8}}
\put(310,60){\vector(0,-1){8}}

\put(305,30){\line(1,0){10}}
\put(310,30){\vector(0,1){8}}
\put(310,30){\vector(0,-1){8}}

\put(290,90){\line(1,0){40}}
\put(310,0){\vector(0,1){15}}
\put(310,90){\vector(0,-1){15}}
\put(345,0){\makebox(0,0){\textcolor{BlueViolet}{$|\!\Psi_{\mathrm{in}}\!\rangle$}}}
\put(345,90){\makebox(0,0){\textcolor{RedViolet}{$\langle\!\Psi_{\mathrm{fin}}\!\mid$}}}
\put(325,68){\makebox(0,0){\textcolor{RedViolet}{$|b_1\rangle$}}}
\put(325,53){\makebox(0,0){\textcolor{BlueViolet}{$\langle\!b_1\!\mid$}}}

\put(325,38){\makebox(0,0){\textcolor{RedViolet}{$|a_1\rangle$}}}
\put(325,23){\makebox(0,0){\textcolor{BlueViolet}{$\langle\!a_1\!\mid$}}}

\put(260,50){\makebox(0,0){time}}
\put(280,0){\makebox(0,0){$t_{\mathrm{in}}$}}
\put(280,90){\makebox(0,0){$t_{\mathrm{fin}}$}}
\put(280,30){\makebox(0,0){$t_1$}}
\put(280,60){\makebox(0,0){$t_2$}}
\put(390,0){\makebox(0,0){\textcolor{BlueViolet}{\small pre-selection}}}
\put(390,60){\makebox(0,0){\small Measurement}}
\put(390,50){\makebox(0,0){\small of $|B\rangle\langle B|$}}

\put(390,30){\makebox(0,0){\small Measurement}}
\put(390,20){\makebox(0,0){\small of $|A\rangle\langle A|$}}

\put(400,90){\makebox(0,0){\textcolor{RedViolet}{\small post-selection}}}
\put(260,40){\vector(0,-1){10}}
\put(260,60){\vector(0,1){10}}

\end{picture}

\caption[One particle superposed in three boxes] 
{\small a) pre-selected vector $|\Psi_{\mathrm{in}} \rangle 
= 1/\sqrt 3 ~(|A\rangle + |B\rangle +|C\rangle)$ propagates
forwards in time from $t_{\mathrm{in}}$ to $t_1$, and post-selected vector $|\Psi_{\mathrm{fin}} \rangle = 1/\sqrt
3   ~(|A\rangle + 
|B\rangle -|C\rangle)$ propagating backwards in time from $t_{\mathrm{fin}}$
to $t_2$.  b)  IM of $\mathrm{\bf \hat{P}}_{A}$ at $t_1$ and of $\mathrm{\bf \hat{P}}_{B}$ at $t_2$.}
\label{suppart}
\end{figure}

Now $Prob(\mathrm{\bf \hat{P}}_{A}=1)=1$ if only box $A$ is opened, while $Prob(\mathrm{\bf \hat{P}}_{\mathrm{B}}=1)=1$ if only box $B$ is opened. When we measure both box $A$ and box $B$, then the particle will  
not be found in both boxes, i.e. $\mathrm{\bf \hat{P}}_{\mathrm{A}}\mathrm{\bf \hat{P}}_{\mathrm{B}}=0$.  But $\mathrm{\bf \hat{P}}_{\mathrm{A}}$ and $\mathrm{\bf \hat{P}}_{\mathrm{B}}$
commute with each other, so how is it possible that measurement of one box can disturb the
measurement of another?
The reason suggested here is that in order to deduce $\mathrm{\bf \hat{P}}_{\mathrm{A}}=1$ from ABL (or $\mathrm{\bf \hat{P}}_{\mathrm{B}}=1$), we used information from both
the pre-and the post-selected vectors (a situation we call diagonal-PPS); when we actually measure  $\mathrm{\bf \hat{P}}_{\mathrm{A}}$, then this measurement will  limit the ``propagation" of the 2-vectors that were relied on to ascertain intermediate values (see fig. \ref{suppart}.b).
If we then subsequently were to measure  $\mathrm{\bf \hat{P}}_{\mathrm{B}}$, the
necessary information from both the pre- and post-selected vectors is no longer available (i.e. information from $t_{\mathrm{in}}$ cannot propagate beyond the measurement of $\mathrm{\bf \hat{P}}_{A}$ at time $t_1$ due to the disturbance caused by an IM).  {\it Thus, even though $\mathrm{\bf \hat{P}}_{\mathrm{A}}$ and $\mathrm{\bf \hat{P}}_{\mathrm{B}}$ commute, measurement of one can disturb measurement of the other with diagonal-PPSs,} and this is related to a violation of the product rule \ref{prodrule}. (In general, if $|\Psi_1\rangle$ is an eigenvector of $\hat{A}$  with eigenvalue $a$ and $|\Psi_2\rangle$ is an eigenvector of $\hat{B}$  with eigenvalue $b$ and $[\hat{A},\hat{B}]=0$, then if $\hat{A}$ and $\hat{B}$ are known only by either pre-selection {\bf or} post-selection, then the product rule is valid, i.e. $\hat{A}\hat{B}=ab$.  However if $\hat{A}$ and $\hat{B}$ are known by both pre-selection {\bf and} post-selection, then the product rule is not valid, i.e. $\hat{A}\hat{B}\neq ab$, i.e. they can still disturb each other, even though they commute.~\cite{vaidman1993b})
\vskip -2cm

\subsection{\bf Weak Measurements}
\label{wmrev}
WMs can be quantified in the quantum measurement theory developed by von Neumann~\cite{vn}:
 to measure an observable $\hat{A}$ of the system, one may use an interaction Hamiltonian of the form 
$H_{\mathrm{int}}=-\lambda  (t)\hat{Q}\hat{A}$
where $\hat{Q}$ is an observable of the measuring device (MD) and $\lambda (t)$ is a coupling constant which is non-zero only during a short time $(0,T)$.  
Using the Heisenberg equations of motion for the momentum $\hat{P}$ of MD (conjugate to the position $\hat{Q}$), we see that $\hat{P}$ changes according to $\frac{d\hat{P}}{d t}=\lambda  (t) \hat{A}$.
Integrating this, we see that $P(T)-P(0)=\lambda A$
where $\int_0^T \lambda  (t)dt=\lambda $.  
To make a more precise determination of $\hat{A}$ requires that either a) $P(0)$ and $P(T)$ are more precisely defined or b) $\lambda$ is large.

A WM can be characterized by either a) $\hat{P}$ of MD is measured to a finite
 precision
 $\Delta P$, (which limits the disturbance  by a
 finite amount $\Delta Q\geq 1/\Delta P$) or b) small $\lambda$.  After the WM interaction, the system is post-selected. 
In this regime, the measurement becomes  less  precise because the uncertainty $\Delta P$ in the position of
 the pointer is larger than the difference in the shifts of
 the pointer $\lambda a_i$ corresponding to the different
 eigenvalues and thus the shift in MD is much smaller than its uncertainty.
The simplest derivation of the WV result is with the second approach, i.e. $\lambda$ small ($\int \lambda  (t)dt=\lambda <<1$): 
\beq
\bra{\Psi_{\mathrm{fin}}}e^{ -i \lambda  \hat{Q} \hat{A}
 }\ket{\Psi_{\mathrm{in}}}\approx\langle\Psi_{\mathrm{fin}}\!\mid
\Psi_{\mathrm{in}} \rangle \lbrace 1+i\lambda \hat{Q} A_{\mathrm{w}})\approx \langle\Psi_{\mathrm{fin}}\ket{\Psi_{\mathrm{in}}}e^{ -i \lambda  \hat{Q}A_{\mathrm{w}}
 }
\eeq
The final
 state
 of MD is almost un entangled with the
 system and is shifted by the WV $A_{\mathrm{w}}$ (assuming without lack of generality that the state of the MD is a Gaussian with spreads $\Delta\equiv\Delta P=\Delta Q=1$): 
\begin{eqnarray}
 \tilde{\Phi}_{\mathrm{fin}}^{\mathrm{MD}}(P) & \to &  \bra{\Psi_{\mathrm{fin}}}e^{ -i \lambda  \hat{Q} \hat{A}
 }\ket{\Psi_{\mathrm{in}}}\tilde{\Phi}_{\mathrm{in}}^{\mathrm{MD}}(P)=\exp\left\{{-{{(P-\lambda \,
 A_{\mathrm{w}})^2}
}}\right\}\\
where \,\,\,A_{\mathrm{w}}&=&\weakv {\Psi_\mathrm{fin}}{\hat{A}}{\Psi_\mathrm{in} }
\label{wv1}
 \label{post_selected}
\end{eqnarray}

\noindent For the PPS cases considered, we do not need to perform the above calculation to obtain the WV.  Instead, we can easily ascertain the WVs due to the following theorems:
\bigskip

\noindent {\bf Theorem 1}: The sum of the WVs is equal to the WV of the sum:
\beq
if \,\,\, (\mathrm{\bf \hat{P}}_{A})_{\mathrm{w}}=(\mathrm{\bf \hat{P}}_{B} +\mathrm{\bf \hat{P}}_{C} )_{\mathrm{w}} \,\,\, then \,\,\, (\mathrm{\bf \hat{P}}_{A})_{\mathrm{w}}=(\mathrm{\bf \hat{P}}_{B})_{\mathrm{w}} +(\mathrm{\bf \hat{P}}_{C})_{\mathrm{w}}
\eeq
\noindent Proof: this follows simply from the linearity of the operators
\beq
{ {\langle \Psi _{\mathrm{fin}} \mid \mathrm{\bf \hat{P}}_{B} + \mathrm{\bf \hat{P}}_{C} \mid \!\Psi _{\mathrm{in}}
\rangle} \over {\langle \Psi _{\mathrm{fin}} \mid \!\Psi _{\mathrm{in}}
\rangle}}= { {\langle \Psi _{\mathrm{fin}} \mid \mathrm{\bf \hat{P}}_{B} \mid \!\Psi _{\mathrm{in}}
\rangle} \over {\langle \Psi _{\mathrm{fin}} \mid \!\Psi _{\mathrm{in}}
\rangle}}+ {{\langle \Psi _{\mathrm{fin}} \mid \mathrm{\bf \hat{P}}_{C} \mid \!\Psi _{\mathrm{in}}
\rangle} \over {\langle \Psi _{\mathrm{fin}} \mid \!\Psi _{\mathrm{in}}
\rangle}}
\label{expweak2}
\eeq

\bigskip

\noindent {\bf Theorem 2}: If a single IM of an observable $\mathrm{\bf \hat{P}}_{A}$ is performed between the pre- and the post-selection, then if the outcome is definite (e.g. $Prob(\mathrm{\bf \hat{P}}_{A}=1$)=1) then the WV is equal to
this eigenvalue (e.g. ($\mathrm{\bf \hat{P}}_{A})_{\mathrm{w}}=1$)~\cite{jmav}. 

\bigskip

\noindent Proof: Given that $\mathrm{\bf \hat{P}}_{A}=\sum_n a_n\ket{\alpha_n}\bra{\alpha_n}$, if an eigenvalue, e.g. $\mathrm{\bf \hat{P}}_{A}=a_n$,  is obtained with certainty, then for $n\neq m$, $\mathrm{\bf \hat{P}}_{A}\equiv \ket{\alpha_m}\bra{\alpha_m}=0$ because the probability to obtain another eigenvalue by ABL is $\propto\la\Psi_{\mathrm{fin}}\ket{\alpha_m}\bra{\alpha_m}\Psi_{\mathrm{in}}\ra=0$.  In this case, the WV $(\mathrm{\bf \hat{P}}_{A})_{\mathrm{w}}=(\ket{\alpha_m}\bra{\alpha_m})_{\mathrm{w}}=\weakv{\Psi_{\mathrm{fin}}}{\, {\alpha_m}\ra\la{\alpha_m} \, }{\Psi_{\mathrm{in}}}=0$. In addition,
\beq
\sum_m\weakv{\Psi_{\mathrm{fin}}}{\, {\alpha_m}\ra\la{\alpha_m} \, }{\Psi_{\mathrm{in}}}=1
\eeq
because $\sum_m \ket{\alpha_m}\bra{\alpha_m}=1$.  But since $\la\Psi_{\mathrm{fin}}\ket{\alpha_m}\bra{\alpha_m}\Psi_{\mathrm{in}}\ra=0$ for $n\neq m$, the only term left is $n$. Therefore, the WV is $1$, the same as the ideal value.

\bigskip
\noindent From Theorem 2, we know the following WVs in the 3-box-paradox with certainty:
\begin{equation}
  \label{psi1}
(\mathrm{\bf \hat{P}}_{\mathrm{A}})_{\mathrm{w}} =1,~~ (\mathrm{\bf \hat{P}}_{\mathrm{B}})_{\mathrm{w}} =1,~~ \mathrm{\bf \hat{P}}_{total}=(\mathrm{\bf \hat{P}}_{\mathrm{A}}+ \mathrm{\bf \hat{P}}_{\mathrm{B}} +\mathrm{\bf \hat{P}}_{\mathrm{C}})_{\mathrm{w}} =1 . 
\end{equation}
Using theorem 1, we obtain:
\begin{eqnarray}
  \label{psi2}
(\mathrm{\bf \hat{P}}_{\mathrm{C}})_{\mathrm{w}}  &=& {\langle\Psi_{\mathrm{fin}}\vert \mathrm{\bf \hat{P}}_{total}-\mathrm{\bf \hat{P}}_{\mathrm{A}}-\mathrm{\bf \hat{P}}_{\mathrm{B}}\vert\Psi_{\mathrm{in}}\rangle
  \over \langle\Psi_{\mathrm{fin}}\vert\Psi_{\mathrm{in}}\rangle}\nonumber\\
&=& (\mathrm{\bf \hat{P}}_{\mathrm{A}}+ \mathrm{\bf \hat{P}}_{\mathrm{B}} +\mathrm{\bf \hat{P}}_{\mathrm{C}})_{\mathrm{w}}- (\mathrm{\bf \hat{P}}_{\mathrm{A}})_{\mathrm{w}} -(\mathrm{\bf \hat{P}}_{\mathrm{B}})_{\mathrm{w}}
= -1 . 
\end{eqnarray}
This surprising theoretical prediction of TSQM has been verified experimentally using photons~\cite{stein1}.
What interpretation should be given to $(\mathrm{\bf \hat{P}}_{\mathrm{C}})_{\mathrm{w}}  =-1$?  One may speculate for formal reasons that this corresponds to a ``negative probability."  However, as will be shown subsequently, this interpretation cannot have any experimental meaning.   On the other hand, we can give it a different interpretation that {\bf does} have an  experimental meaning if we perform any WM which is sensitive to the projection operator $\mathrm{\bf \hat{P}}_{\mathrm{C}}$.  In this case, we will observe the opposite effect from those cases in which the projection operator is positive.  This suggests that there is $-1$ particle in box $C$, e.g. a WM of the amount of charge in  box $C$ in
the intermediate 
time will yield a negative charge (assuming it is a positively charged particle).

\bigskip

\subsection{\bf Contextuality and PPS-paradoxes}
In \cite{jt} it was first pointed out and extensively discussed and later proven by Leifer and Spekkens~\cite{sl2}, that whenever there is a logical-PPS-paradox (as in the 3-box-paradox), 
then there is a related proof of contextuality.  
The proof considers ``all the measurements defined by the PPS-paradox --the preselection, the post-selection, and the alternative possible intermediate measurements--as alternative possible measurements at a single time."
They show a direct connection  between each of the 8 vectors in the 3-box-paradox and the 8 vectors in the Clifford/BKS-proof: it is readily seen (\cite{sl2}'s figure)
 that no HVT assignment of $0's$ (black circles) and $1's$ (white circles) can be made that is consistent with the orthogonality relations, since in order for non-contextuality to hold, no 2 orthogonal pairs can be white, 
 which is violated by the central 2 circles.

On the one hand, we will show how definite values can be assigned to observables (verified by IMs on PPSs) in surprising ways.  While this  assignment  suggests novel connections between  what could be said about the state before the IM and after, in general, the IM creates a disturbance and thus creates an uncertain relationship between the state before and after, reflecting Bub and Brown's~\cite{bub1} concern ``ensembles which have been preselected and post-selected via an arbitrary intervening measurement...are not well defined without specification of the intervening measurement."  However, this is not the case for WMs (PPS ensembles are well-defined for any intermediate WM).  
In addition, we show~\cite{jt} how measurement disturbance can arise in new ways for IMs of commuting observables in PPS situations, and so it was also argued~\cite{jt,sl}  that a noncontextual-HVT can reproduce QM if we allow for a disturbance of the HVs: ``the possibility of measurement disturbance blocks the conclusion that a PPS-paradox is itself a proof of the contextuality of HVTs."  
The motivation~\cite{sl} behind this assertion was the belief that PPS-paradoxes could be explained entirely within CM~\cite{kastner,kirkpatrick} and  that contextuality should not be regarded as fundamental in a classical picture of reality.  
Nevertheless, the ``paradoxical," i.e. non-classical, nature of the 3-box-paradox was recently re-affirmed in terms of IMs~\cite{lev06}.  In this article, we expand this point mainly by showing how non-classical WVs can be empirically demonstrated in PPS-paradoxes. 
E.g. WMs on  box $C$ will record the ``paradoxical," non-classical outcome of  $-1$.
WVs like this outside the eigenvalue spectrum cannot be reproduced from any positive definite probability distribution of eigenvalues~\cite{at3, ab, jt}. 

Finally, with WMs, there is no measurement disturbance, yet non-classical results are still obtained.  In \S \ref{ppscontext} we show how WMs allow us to make a stronger connection between PPSs and contextuality.  This weakens the assertion that~\cite{sl} ``PPS paradoxes do not require contextuality for their explanation but do require measurement-disturbance."

\subsubsection{\bf HVTs and PPSs: }
\label{hvtpps}
In~\cite{sl}, quantum states are given as probability distributions $\mu$ over HVTs $\lambda$, such that $\int_{\Omega }\mu (\lambda )d\lambda
=1$, where $\Omega$ is the set of possible HVTs and   
measurements are characterized by 
idempotent indicator
functions $\chi ^{\mathrm{M}}_{j}:\Omega \rightarrow \{0,1\}$, such that $\sum_{j}\chi ^{\mathrm{M}}_{j}(\lambda
)=1$.
The probability to obtain an outcome $j$ for a random variable $X$ when a measurement $M$ is performed on a system in $\mu$ then is  
\begin{equation}
p_{\mu}(X_{\mathrm{M}}=j)=\int_{\Omega }\chi^\mathrm{M}_j(\lambda )\mu (\lambda )d\lambda   \label{HVTProb}
\end{equation}
In addition, a transition from $\omega $ to some other HVT state $\lambda $ could result from the measurement process~\cite{sl}, and this is modeled  by a transition probability 
$D^{\mathrm{M}}_{j}(\lambda ,\omega )$ such that  $\int_{\Omega }D^\mathrm{M}_j(\lambda ,\omega )d\lambda =1.$ 
This approach to HVTs may then be applied~\cite{sl2} to ABL:
\begin{equation}
p_{\mathrm{HVT}}(X_{\mathrm{M}}=k|\mathrm{A}_{\mathrm{pre}},\mathrm{A}_{\mathrm{post}},
\mathrm{M})=\frac{\int_{\Omega }\chi _{\mathrm{post}}(\lambda )\Gamma_{k}^{\mathrm{M}
}(\lambda ,\omega )\mu _{\mathrm{pre}}(\omega )d\omega d\lambda
}{\int_{\Omega }\chi _{\mathrm{post}}(\lambda
)(\Gamma_{k}^{\mathrm{M}}(\lambda ,\omega )+\Gamma_{\lnot
k}^{\mathrm{M}}(\lambda ,\omega ))\mu _{\mathrm{pre}}(\omega )d\omega
d\lambda } \label{ablhvt}
\end{equation}

\noindent WMs and WVs can also be described in this language (the usual projective measurement typically utilized in quantum experiments are a special case of WMs~\cite{brun}) and this will be developed in a subsequent article~\cite{at6}. Some general considerations on possible relationships between  HVTs and WVs are:
\begin{itemize}
\item {\bf Epistemic nature of probability: }
If the probabilistic nature of 
$D^{\mathrm{M}}_{j}(\lambda ,\omega )$ and $\chi^\mathrm{M}_j(\lambda )$ are of epistemic origin, then this HVT approach~\cite{sl} requires that non-commuting observables (such as $p$ and $x$) can have a ``simultaneous" precise reality, as suggested, e.g. by the Wigner-Moyal method.  
If we require that any 
theoretical formalism should include exactly what can be measured (no more and no less),  
 then it should be possible to make measurements on these projections.
While such densities do give the correct average of a function, i.e. $\int \rho (x,p) f(x,p) dxdp$ (thus  appear to behave as proper densities),  they also have un-physical aspects, i.e. mathematical artefacts, when the densities become negative.  The reason (as will be shown subsequently~\cite{at6}) is that if we attempt to actually measure such ``negative" properties, then the result does not correspond to a physical observable  in Hilbert Space.  E.g. if we did try to project on $p$ and $x$ as densities simultaneously, then we obtain the parity operator, taking a generic $\psi (x)$ to $\psi (-x)$.  To see this, we translate the classical projection $p=0$ and $x=0$ into QM:
\beq
\int_{-\infty}^{\infty} \int_{-\infty}^{\infty} e^{i\alpha x+i\beta p}d\alpha d\beta
\underbrace{\Rightarrow}_{QM}\int_{-\infty}^{\infty} \int_{-\infty}^{\infty} e^{\frac{i\alpha\beta}{2}}e^{i\alpha \hat{x}}e^{i\beta \hat{p}}d\alpha d\beta
\eeq
Consider applying this to a generic wavefunction. First, the exponential, $e^{i\beta \hat{p}}$, translates $\psi (x)$.  Integrating then over $\alpha$ produces a delta function:
\beq
\int_{-\infty}^{\infty} \int_{-\infty}^{\infty}  e^{i\frac{\alpha\beta}{2}} e^{i\alpha x}\psi (x+\beta) d\alpha d\beta =\int_{-\infty}^{\infty} \left\{\underbrace{\int_{-\infty}^{\infty} e^{i\alpha (x+\frac{\beta}{2})}  d\alpha}_{\delta(x+\frac{\beta}{2})} \right\} \psi (x+\beta) d\beta
\eeq

Finally, integrating over $\beta$, we obtain $\beta=-2x$, and thus $\psi (x-2x)=\psi (-x)$.  Therefore, the quantum analog of the classical projection does not correspond to a quantum projector: it corresponds to a highly non-local result, the parity operator.

\item {\bf Ontological nature of probability: }
In contrast to this non-physical aspect,  non-classical  WVs (e.g.  negative WVs for projection operators)  {\bf can} be seen experimentally.  There are additional advantages besides this greater harmony between measurements and theory~\cite{at6}.  E.g. if the quantum/probabilistic nature of $D^{\mathrm{M}}_{j}(\lambda ,\omega )$ is taken to be fundamental, 
then 
a natural/axiomatic explanation can be obtained~\cite{at6} for 2 seemingly opposite situations: a) the contextual dependence of value assignments on details of MD and b) the fact that the statistics do not depend on the details of MD.  
These can be harmonized as a consistency check with causality~\cite{at6} as indicated by inserting a complete set of states $\{ \ket{\Psi_{\mathrm{fin}}}_j \}$ into 
$\bar{A}$:
\beq
 \bar{A} =  \bra{\Psi_{\mathrm{in}}} { \left[\sum_j  \ket{\Psi_{\mathrm{fin}}}_j\bra{\Psi_{\mathrm{fin}}}_j\right]\hat{A}} \ket{\Psi_{\mathrm{in}}}
= \sum_j |\langle \Psi _{\mathrm{fin}} \!\mid_j \!\Psi _{\mathrm{in}}
\rangle|^2\ 
{ {\langle \Psi _{\mathrm{fin}}\! \mid_j \hat{A} \mid \!\Psi _{\mathrm{in}}
\rangle} \over {\langle \Psi _{\mathrm{fin}} \!\mid_j \!\Psi _{\mathrm{in}}
\rangle}}
\label{expweak}
\eeq
Thus, one can think of  $\bar{A}$ for the whole ensemble as being built out of pre- and post-selected
states in which   the WV is multiplied by a probability for post-selection.
The fluctuation in the system then is also relevant for the probability to obtain different  post-selections: as the fluctuation in the system 
increases, the probability of a rare or eccentric post-selection also increases.  However an attempt  to see this through WMs will require 
the spread in the MD to be increased  and this increases the probability of seeing
the strange result as an error of the MD.

\item {\bf Time-symmetry versus time-asymmetry: }
The HVT approach utilized in~\cite{sl} focuses on the probability of post-selection and brings in an element of time-asymmetry, apparently endemic to a Bayesian approach~\cite{ab}.
\end{itemize}
\bigskip

\subsubsection{\bf Logical-PPS-paradoxes imply  contextuality through WMs: }
\label{ppscontext}
A logical-PPS-paradox occurs if $p_{\mathrm{HVT}}=1$ for several incompatible situations.  This can occur if 
\beq
{\int_{\Omega }\chi _{\mathrm{post}}(\lambda
)\Gamma_{\lnot
k}^{\mathrm{M}}(\lambda ,\omega )\mu _{\mathrm{pre}}(\omega )d\omega
d\lambda }=0
\eeq
However, 
Leifer and Spekkens~\cite{sl} argue that the elements in the PPS-paradox have nothing to do with contextuality  because their interpretation is different: ``..we did not show that a logical PPS paradox is {\it itself}  a proof of contextuality; measurements that are temporal  successors in the PPS paradox must be treated as counterfactual alternatives in the proof of contextuality.  This distinction is critical, since an earlier measurement can cause a disturbance to the ontic state that is monitored by a later measurement".
However, WMs (instead of IMs) can give an empirical meaning to ``alternative possible measurements at a single time".  Although these alternatives are usually regarded as counter-factual, 
they will all be true simultaneously with WMs. 

\bigskip

\noindent {\bf Theorem:} Logical-PPS-paradoxes imply contextuality through WMs. 
Proof: 
Theorem 2 allows us to state that  
all counterfactual statements which maintain  the occurence of an outcome with
certainty will all be true simultaneously when they are measured weakly.   Theorem 2 is applicable to the precise elements utilized in the contextuality proof~\cite{sl2}.  In addition, given that WVs are by definition independent of the type of WM and given eq. \ref{expweak}, and since WVs can violate the algebraic conditions (the product rule) required for BKS and noncontextual-HVTs~\cite{sl}, we have now proven that logical-PPS-paradoxes which assign definite probabilities (of $0$ or $1$) via ABL, are in fact proofs of contextuality if all the ``alternative possible intermediate measurements" are performed weakly and the argument that their interpretation is different does not apply.

We have thereby mitigated the attempt to explain these PPS  ``paradoxes" as a result of disturbance (since there is no disturbance with WMs)
  and have strengthened~\cite{sl2} the connection between these ``paradoxes" (which we will argue cannot be reproduced by a ``classical-like" HVT) and contextuality.

\section{\bf PPS AND CONTEXTUALITY IN 4-D}
In the 3-box-paradox, the product of observables was always definite, i.e. $\mathrm{\bf \hat{P}}_{\mathrm{A}}\mathrm{\bf \hat{P}}_{\mathrm{B}}=0$ and the proof of contextuality was state dependent.  In this section, we consider a slightly  different situation (4D BKS nonets) in which the product of observables can give 2 different values.   Except for this difference, this 4D BKS nonet example is similar to the 3-box-paradox in that we shall also 
analyze them in terms of PPSs thereby revealing surprising predictions for IMs and WMs and will  
demonstrate the identical issues of diagonal-PPS measurements, violation of the product rule, contextuality, and WMs which cannot be explained by a noncontextual-HVT.

\subsection{\bf Review of Mermin's 4-D BKS theorem}
We consider Mermin's  version of BKS with a set of 9 observables.  It is intuitive~\cite{merminKS} to represent all the 
``functional relationships between mutually commuting subsets of the observables," i.e.  $V_{\vec{\psi}}(F\{\hat{A}\})=F\{V_{\vec{\psi}}(\hat{A})\}$,
by drawing them in fig. 2 and arranging them so that all the observables
in each row (and column) commute with all the other observables in the same row (or column). 
\vskip -1cm
\singlespacing
\singlespacing
\begin{figure}[h] 
\begin{center}
\begin{eqnarray*}
& {\hat{\sigma}_x}^1  \;\;\;  \;\;\;\;\;\;\;\;\;\;\;\; {\hat{\sigma}_x}^2 
 \;\;\; \;\;\;\;\;\;\; {\hat{\sigma}_x}^1{\hat{\sigma}_x}^2 \;\; \;\;\;=1\nonumber \\
\nonumber \\
& {\hat{\sigma}_y}^2 \;\;\; \;\;\;\;\;\;\;\;\;\;\;\; {\hat{\sigma}_y}^1 
 \;\;\; \;\;\;\;\;\;\; {\hat{\sigma}_y}^1{\hat{\sigma}_y}^2 \;\; \;\;\;=1\nonumber \\ 
\nonumber \\
& {\hat{\sigma}_x}^1{\hat{\sigma}_y}^2\;\;\;  \;\;\;\;\; {\hat{\sigma}_x}^2 {\hat{\sigma}_y}^1
\;\;\; \;\; \;\;\; {\hat{\sigma}_z}^1{\hat{\sigma}_z}^2\;\; \;\;\;=1 \nonumber \\ 
\nonumber \\
& \;\;=1 \;\; \;\;\;\;\;\; =1 \;\; \;\;\;\;\; =-1 \;\;\;\;\;\; \;\;\;\;\;\nonumber \\ 
\end{eqnarray*}
\end{center}
\begin{picture}(400,90)(0,0)
\color{BlueViolet}

\put(210,159){\oval(180,20)}
\put(210,195){\oval(180,20)}
\put(210,235){\oval(180,20)}

\put(141,179){\oval(39,135)}
\put(196,179){\oval(39,135)}

\color{Red}
\put(251,179){\oval(39,135)}

\end{picture}
\vskip -10pc

\label{merm4dov}
\caption[]{4-D BKS example}

\end{figure}
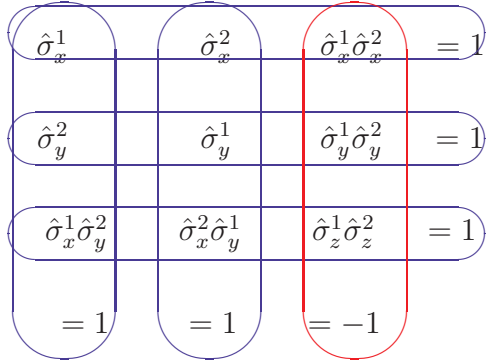

Individually, each of the 9 observables depicted in fig. 2
has eigenvalues $\pm 1$. 
A noncontextual-HVT requires an assignment of values that is $\pm 1$.  
In addition, eq. \ref{KS1} requires that 
the value assigned to the product of all three observables in any row or column must obey the same identities that the observables themselves satisfy, i.e. the product of the values assigned to the observables in each oval yields a result of $+1$ except in the last column which gives $-1$.  
(The value assignments are given by $V_{\vec{\psi}}({\hat{\sigma}_x}^1 )=\langle \hat{\sigma}^1_x\bigotimes I^2\rangle$, $V_{\vec{\psi}}({\hat{\sigma}_x}^2 )=\langle I^1\bigotimes \hat{\sigma}^2_x\rangle$... $V_{\vec{\psi}}({\hat{\sigma}_x}^1 )=\langle \hat{\sigma}^1_z\bigotimes \hat{\sigma}^2_z\rangle$ 
). Computing column  3 of fig. 2: 
\begin{eqnarray}
\{\hat{\sigma}^1_x\hat{\sigma}^2_x\}\{\hat{\sigma}^1_y\hat{\sigma}^2_y\}\{\hat{\sigma}^1_z\hat{\sigma}^2_z\}&=&\hat{\sigma}^1_x\underbrace{\hat{\sigma}^2_x\hat{\sigma}^1_y}_{commute\,so\,\hookrightarrow}\hat{\sigma}^2_y\hat{\sigma}^1_z\hat{\sigma}^2_z=
\underbrace{\hat{\sigma}^1_x\hat{\sigma}^1_y}_{=i\hat{\sigma}^1_z}\underbrace{\hat{\sigma}^2_x\hat{\sigma}^2_y}_{=i\hat{\sigma}^2_z}\hat{\sigma}^1_z\hat{\sigma}^2_z\nonumber\\
&=&
i\hat{\sigma}^1_z\underbrace{i\hat{\sigma}^2_z\hat{\sigma}^1_z}_{commute\,so\,\hookrightarrow}\hat{\sigma}^2_z= i\hat{\sigma}^1_zi\hat{\sigma}^1_z\hat{\sigma}^2_z\hat{\sigma}^2_z= -1
\end{eqnarray}
Computing the product of the observables in the third row, i.e.:
\begin{eqnarray}
\{\hat{\sigma}^1_x\hat{\sigma}^2_y\}\{\hat{\sigma}^2_x\hat{\sigma}^1_y\}\{\hat{\sigma}^1_z\hat{\sigma}^2_z\}&=&\hat{\sigma}^1_x\underbrace{\hat{\sigma}^2_y\hat{\sigma}^2_x}_{=-i\hat{\sigma}^2_z}\hat{\sigma}^1_y\{\hat{\sigma}^1_z\hat{\sigma}^2_z\}=\underbrace{\hat{\sigma}^1_x\hat{\sigma}^1_y}_{=i\hat{\sigma}^1_z}\underbrace{\{-i\hat{\sigma}^2_z\}\{\hat{\sigma}
^1_z}_{commute\,so\,\hookrightarrow}\hat{\sigma}^2_z\}\nonumber\\
&=&\underbrace{i\hat{\sigma}^1_z \hat{\sigma}^1_z}_{=i}
\underbrace{\{-i\hat{\sigma}^2_z\}\{\hat{\sigma}^2_z\}}_{=-i}=+1,
\end{eqnarray}
If the product rule  is applied to the value assignments made in the rows, then:
\begin{eqnarray}
V_{\vec{\psi}}({\hat{\sigma}_x}^1 )V_{\vec{\psi}}({\hat{\sigma}_x}^2 )V_{\vec{\psi}}({\hat{\sigma}_x}^1{\hat{\sigma}_x}^2)=
V_{\vec{\psi}}({\hat{\sigma}_y}^2)V_{\vec{\psi}}({\hat{\sigma}_y}^1 )V_{\vec{\psi}}( {\hat{\sigma}_y}^1{\hat{\sigma}_y}^2)&=&\nonumber\\
V_{\vec{\psi}}({\hat{\sigma}_x}^1{\hat{\sigma}_y}^2)V_{\vec{\psi}}({\hat{\sigma}_x}^2 {\hat{\sigma}_y}^1)V_{\vec{\psi}}({\hat{\sigma}_z}^1{\hat{\sigma}_z}^2)&=&+1
\label{rows}
\end{eqnarray}
while the column identities require:
\begin{eqnarray}
V_{\vec{\psi}}({\hat{\sigma}_x}^1 )V_{\vec{\psi}}({\hat{\sigma}_y}^2)V_{\vec{\psi}}({\hat{\sigma}_x}^1{\hat{\sigma}_y}^2)=
V_{\vec{\psi}}({\hat{\sigma}_x}^2 )V_{\vec{\psi}}({\hat{\sigma}_y}^1 )V_{\vec{\psi}}({\hat{\sigma}_x}^2 {\hat{\sigma}_y}^1)&=&+1\nonumber\\
V_{\vec{\psi}}({\hat{\sigma}_x}^1{\hat{\sigma}_x}^2)V_{\vec{\psi}}( {\hat{\sigma}_y}^1{\hat{\sigma}_y}^2)V_{\vec{\psi}}({\hat{\sigma}_z}^1{\hat{\sigma}_z}^2)&=&-1
\label{columns}
\end{eqnarray}
However, it is easy to show that the 9 numbers $V_{\vec{\psi}}$ cannot satisfy all 6 constraints:  multiplying all $9$ observables together gives 2 different results, a $+1$ when it is done row by row and a $-1$  when it is done column by column: 

\begin{equation}
V_{\vec{\psi}}({\hat{\sigma}_x}^1 )V_{\vec{\psi}}({\hat{\sigma}_x}^2 )V_{\vec{\psi}}({\hat{\sigma}_x}^1{\hat{\sigma}_x}^2)...V_{\vec{\psi}}({\hat{\sigma}_z}^1{\hat{\sigma}_z}^2)=+1
\label{1sthvt}
\end{equation}
\begin{equation}
V_{\vec{\psi}}({\hat{\sigma}_x}^1 )V_{\vec{\psi}}({\hat{\sigma}_y}^2)V_{\vec{\psi}}({\hat{\sigma}_x}^1{\hat{\sigma}_y}^2)...V_{\vec{\psi}}({\hat{\sigma}_z}^1{\hat{\sigma}_z}^2)=-1
\label{2ndhvt}
\end{equation}
There obviously is no consistent solution to eqs. \ref{2ndhvt} and \ref{1sthvt} since they contain the same set of numbers, simply ordered differently.
Therefore the values assigned to the observables cannot obey the same identities that the observables themselves obey, $V_{\vec{\psi}}(F\{\hat{A}\})\neq F\{V_{\vec{\psi}}(\hat{A})\}$, 
 and an HVT would have to assign  values to observables in a
way that depended on the choice of which of 2 mutually commuting sets of observables that were
also chosen to measure, i.e. the values assigned are contextual.

\subsection{\bf ABL, VAA and BKS nonets}
\label{vaabks}
Following Vaidman, Albert, and Aharonov (VAA)~\cite{s-xyz}, Mermin showed how to assign a definite value to a single measurement of any one of the nine observables of a BKS nonet ~\cite{mermin2v}.
He then generalized this to a definite assignment to any one of 16 observables from the sets of nonets  and showed that this assignment cannot be done if one attempts to measure (or ascertain) 2 or more of the observables belonging to the nonets. He left open the question as to the physical reason for this, stating ``I find this intriguing"~\cite{mermin2v}  To address this, we present a physical reason to demonstrate why the VAA scheme cannot be applied to 2 or more measurements by showing that the 2 measurements interfere with each other given the necessary PPSs.  
This can be seen to be a consequence of TSQM: some assignment of eigenvalues to operators are based on just one of the two vectors (i.e. either the pre- {\it or} the post-selected vector) while some assignment of eigenvalues are based on both vectors (i.e. both the pre-selected {\it and} the post-selected vectors - what we call diagonal-PPS).  
In this picture, it is the utilization of more than one PPS and the subsequent interference between them that explains the violation of the product rule and thus the physical source of the ``contextuality" (provided
that the contradictory statements are obtained by measuring the product of observables, rather
than  by measuring the observables individually, and this is always the case).
When assignments are not made in the diagonal-PPS sense, then sets of commuting observables which are determined entirely by just one vector {\bf satisfy} the BKS function condition $V_{\vec{\psi}}(F\{\hat{A}\})=F\{V_{\vec{\psi}}(\hat{A})\}$
Sets of commuting observables which are assigned values in the diagonal-PPS sense by using information from both vectors {\bf do not} satisfy the BKS function condition 
because they violate the product rule, 
and can disturb each other.
\vskip 1cm
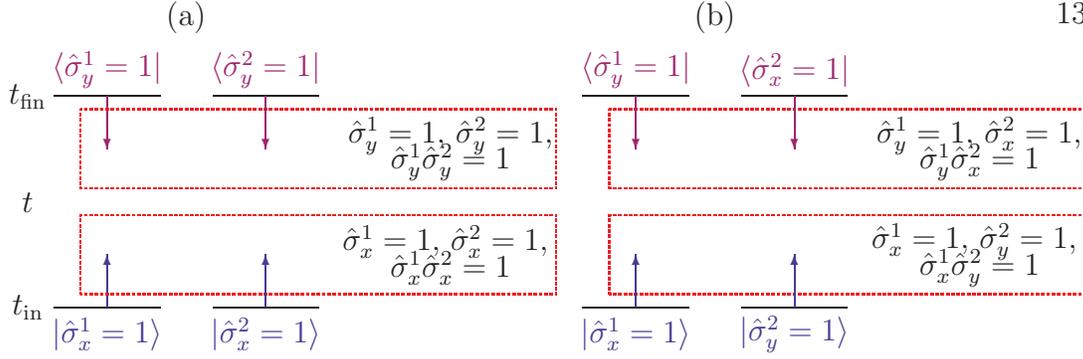
\begin{figure}[h]
 \epsfxsize=4.5truein
\begin{picture}(400,90)(0,0)
\put(40,10){\line(1,0){40}}
\put(40,90){\line(1,0){40}}
\color{BlueViolet}
\put(60,10){\vector(0,1){20}}
\color{RedViolet}
\put(60,90){\vector(0,-1){20}}
\color{Red}
\put(50,55){\dashbox{1}(180,30)}
\put(50,15){\dashbox{1}(180,30)}
\put(250,55){\dashbox{1}(180,30)}
\put(250,15){\dashbox{1}(180,30)}
\color{OliveGreen}

\color{Black}
\put(100,10){\line(1,0){40}}
\put(100,90){\line(1,0){40}}
\color{BlueViolet}
\put(120,10){\vector(0,1){20}}
\color{RedViolet}
\put(120,90){\vector(0,-1){20}}
\color{Black}
\put(240,10){\line(1,0){40}}
\put(240,90){\line(1,0){40}}
\color{BlueViolet}

\put(260,10){\vector(0,1){20}}
\color{RedViolet}

\put(260,90){\vector(0,-1){20}}
\color{OliveGreen}

\color{BlueViolet}
\put(260,0){\makebox(0,0){$|\hat{\sigma}_x^1=1\ra$}}
\put(120,0){\makebox(0,0){$|\hat{\sigma}_x^2=1\ra$}}
\put(60,0){\makebox(0,0){$|\hat{\sigma}_x^1=1\ra$}}
\put(320,0){\makebox(0,0){$|\hat{\sigma}_y^2=1\ra$}}

\color{RedViolet}
\put(260,100){\makebox(0,0){$\la\hat{\sigma}_y^1=1|$}}
\put(120,100){\makebox(0,0){$\la\hat{\sigma}_y^2=1|$}}
\put(60,100){\makebox(0,0){$\la\hat{\sigma}_y^1=1|$}}
\put(320,100){\makebox(0,0){$\la\hat{\sigma}_x^2=1|$}}
\color{Black}
\put(300,10){\line(1,0){40}}
\put(300,90){\line(1,0){40}}
\color{BlueViolet}

\put(320,10){\vector(0,1){20}}
\color{RedViolet}

\put(320,90){\vector(0,-1){20}}
\color{Black}
\put(30,10){\makebox(0,0){$t_{\mathrm{in}}$}}
\put(30,90){\makebox(0,0){$t_{\mathrm{fin}}$}}
\put(30,50){\makebox(0,0){$t$}}
\put(90,120){\makebox(0,0){(a)}}
\put(290,120){\makebox(0,0){(b)}}
\put(190,35){\makebox(0,0){$\hat{\sigma}^1_x=1$, $\hat{\sigma}^2_x=1$, }}
\put(190,25){\makebox(0,0){$\hat{\sigma}^1_x\hat{\sigma}^2_x=1$}}
\put(190,75){\makebox(0,0){$\hat{\sigma}^1_y=1$, $\hat{\sigma}^2_y=1$,}}
\put(190,65){\makebox(0,0){$\hat{\sigma}^1_y\hat{\sigma}^2_y=1$}}

\put(390,35){\makebox(0,0){$\hat{\sigma}^1_x=1$, $\hat{\sigma}^2_y=1$, }}
\put(390,25){\makebox(0,0){$\hat{\sigma}^1_x\hat{\sigma}^2_y=1$}}
\put(390,75){\makebox(0,0){$\hat{\sigma}^1_y=1$, $\hat{\sigma}^2_x=1$,}}
\put(390,65){\makebox(0,0){$\hat{\sigma}^1_y\hat{\sigma}^2_x=1$}}

\color{OliveGreen}

\end{picture}

\caption{\small PPS states for Mermin example.}
\label{2vmermin}
\end{figure}

\subsubsection{\bf Ascertaining the results of any one of the 9 observables: }
\label{asc9}

   We begin our analysis  by considering specific examples of PPS configurations.  We then utilize ABL~\cite{abl} and 
show that choosing different post-selections change which triplet of observables violate the product rule.
Consider first a pre-selection of $\hat{\sigma}^1_x=1$ and $\hat{\sigma}^2_x=1$ and a post-selection of
$\hat{\sigma}^1_y=1$ and $\hat{\sigma}^2_y=1$ (see fig. \ref{2vmermin}.a).  
In this case, 
it is easy to see that we can ascertain with certainty any one of the following values $\hat{\sigma}^1_x=1$,
$\hat{\sigma}^2_x=1$, $\hat{\sigma}^1_y=1$ or $\hat{\sigma}^2_y=1$.  We also know that we will obtain definite
values of $+1$ if we measure any one of the following products of observables: $\hat{\sigma}^1_x\hat{\sigma}^2_x=1$ and 
$\hat{\sigma}^1_y\hat{\sigma}^2_y=1$, so we must also obtain $\hat{\sigma}^1_x\hat{\sigma}^2_x\hat{\sigma}^1_y\hat{\sigma}^2_y =+1$. 
\beq
  \label{psi2}
\langle \hat{\sigma}^1_x\hat{\sigma}^2_x\hat{\sigma}^1_y\hat{\sigma}^2_y
\rangle = \langle\hat{\sigma}^1_y=1\vert\langle \hat{\sigma}^2_y=1\vert \hat{\sigma}^1_x\hat{\sigma}^2_x\hat{\sigma}^1_y\hat{\sigma}^2_y
\rangle\vert\hat{\sigma}^1_x=1\rangle\vert\hat{\sigma}^2_x=1\rangle=+1
\eeq
In addition, we obtain the same results if we switch the sequence of $\hat{\sigma}^1_x\hat{\sigma}^2_x$ and $\hat{\sigma}^1_y\hat{\sigma}^2_y$ because they commute.
Any {\it one} of the other observables in fig. 2 (i.e. $\hat{\sigma}^1_x\hat{\sigma}^2_y$ and $\hat{\sigma}^2_x\hat{\sigma}^1_y$) can also be ascertained with certainty given this PPS. 
Finally, from the product of the 3 observables in column 3 
($\{\hat{\sigma}^1_x\hat{\sigma}^2_x\}\{\hat{\sigma}^1_y\hat{\sigma}^2_y\}\{\hat{\sigma}^1_z\hat{\sigma}^2_z\}=-1$ 
and $\{\hat{\sigma}^1_x\hat{\sigma}^2_x\}\{\hat{\sigma}^1_y\hat{\sigma}^2_y\}=+1$), 
we can deduce that $\hat{\sigma}^1_z\hat{\sigma}^2_z=-1$.
Similar statements can be made for other PPSs. 
E.g. consider the  pre-selection: $\hat{\sigma}^1_x=1$ and $\hat{\sigma}^2_y=1$  and a post-selection of
$\hat{\sigma}^1_y=1$ and $\hat{\sigma}^2_x=1$ (see fig. \ref{2vmermin}.b)  In this case, the  measurement 
$\hat{\sigma}^1_x\hat{\sigma}^2_y\hat{\sigma}^2_x\hat{\sigma}^1_y=+1$ and we can deduce from the third row that $\{\hat{\sigma}^1_z\hat{\sigma}^2
_z\}=+1$. 

Thus, given just one PPS, any single observable can be assigned a definite value, even though $\hat{\sigma}^1_z\hat{\sigma}^2_z$ is assigned different values in different PPS.  
It is precisely because of this connection between particular PPSs and different values for $\hat{\sigma}^1_z\hat{\sigma}^2_z$
that the issue of contextuality arises when we consider products of these observables.

\subsubsection{\bf Ascertaining the results of {\it products} of the 9 observables: }
\label{ascprod}
In this section, we ask how many of the {\it products} of the 9 observables 
in fig. 2 can be ascertained together with certainty.
For example, as stated in the previous section, the outcome for the product of the first two observables in  column 3 of fig. 2
with the PPS of fig. \ref{2vmermin}.a is $\sigma^1_x\sigma^2_x\sigma^1_y\sigma^2_y =+1$.
However, if we measure the operators corresponding to the first 2 observables of row 3 in fig. 2,
i.e.
$\hat{\sigma}^1_x\hat{\sigma}^2_y\hat{\sigma}^2_x\hat{\sigma}^1_y$,
given this particular PPS shown in fig. \ref{2vmermin}.a, then the sequence of measurements interfere with each other (as represented by the slanted ovals in
figure \ref{2vmermindiag}.a).
\begin{figure}[h] \epsfxsize=4.5truein
\vskip 1cm
\begin{picture}(400,90)(0,0)
\put(40,10){\line(1,0){40}}
\put(40,90){\line(1,0){40}}
\color{BlueViolet}
\put(60,10){\vector(0,1){10}}
\color{RedViolet}
\put(60,90){\vector(0,-1){10}}

\color{Black}
\put(100,10){\line(1,0){40}}
\put(100,90){\line(1,0){40}}
\color{BlueViolet}

\put(120,10){\vector(0,1){25}}
\color{RedViolet}

\put(120,90){\vector(0,-1){25}}
\color{OliveGreen}

\color{Black}
\put(240,10){\line(1,0){40}}
\put(240,90){\line(1,0){40}}
\color{BlueViolet}

\put(260,10){\vector(0,1){10}}
\color{RedViolet}

\put(260,90){\vector(0,-1){25}}
\color{OliveGreen}

\color{BlueViolet}
\put(260,0){\makebox(0,0){$|\hat{\sigma}_x^1=1\ra$}}
\put(120,0){\makebox(0,0){$|\hat{\sigma}_x^2=1\ra$}}
\put(60,0){\makebox(0,0){$|\hat{\sigma}_x^1=1\ra$}}
\put(320,0){\makebox(0,0){$|\hat{\sigma}_y^2=1\ra$}}

\color{RedViolet}
\put(260,100){\makebox(0,0){$\la\hat{\sigma}_y^1=1|$}}
\put(120,100){\makebox(0,0){$\la\hat{\sigma}_y^2=1|$}}
\put(60,100){\makebox(0,0){$\la\hat{\sigma}_y^1=1|$}}
\put(320,100){\makebox(0,0){$\la\hat{\sigma}_x^2=1|$}}
\color{Black}
\put(300,10){\line(1,0){40}}
\put(300,90){\line(1,0){40}}
\color{BlueViolet}

\put(320,10){\vector(0,1){25}}
\color{RedViolet}

\put(320,90){\vector(0,-1){10}}
\color{Black}
\put(30,10){\makebox(0,0){$t_{\mathrm{in}}$}}
\put(30,90){\makebox(0,0){$t_{\mathrm{fin}}$}}
\put(30,50){\makebox(0,0){$t$}}
\put(90,120){\makebox(0,0){(a)}}
\put(290,120){\makebox(0,0){(b)}}

\put(80,20){\makebox(0,0){$\hat{\sigma}^1_x=1$}}
\put(110,40){\makebox(0,0){$\hat{\sigma}^2_y=?$}}
\put(110,60){\makebox(0,0){$\hat{\sigma}^2_x=?$}} 
\put(80,80){\makebox(0,0){$\hat{\sigma}^1_y=1$}}

\put(150,20){\makebox(0,0){$t_1$}}
\put(150,40){\makebox(0,0){$t_2$}}
\put(150,60){\makebox(0,0){$t_3$}} 
\put(150,80){\makebox(0,0){$t_4$}}

\put(280,20){\makebox(0,0){$\hat{\sigma}^1_x=1$}}
\put(340,40){\makebox(0,0){$\hat{\sigma}^2_x=?$}}
\put(280,60){\makebox(0,0){$\hat{\sigma}^1_y=1$}} 
\put(340,80){\makebox(0,0){$\hat{\sigma}^2_y=?$}}

\put(370,20){\makebox(0,0){$t_1$}}
\put(370,40){\makebox(0,0){$t_2$}}
\put(370,60){\makebox(0,0){$t_3$}} 
\put(370,80){\makebox(0,0){$t_4$}}

\end{picture}

\caption{\small Time sequence of PPS measurements for Mermin example.}
\label{2vmermin2}
\end{figure}
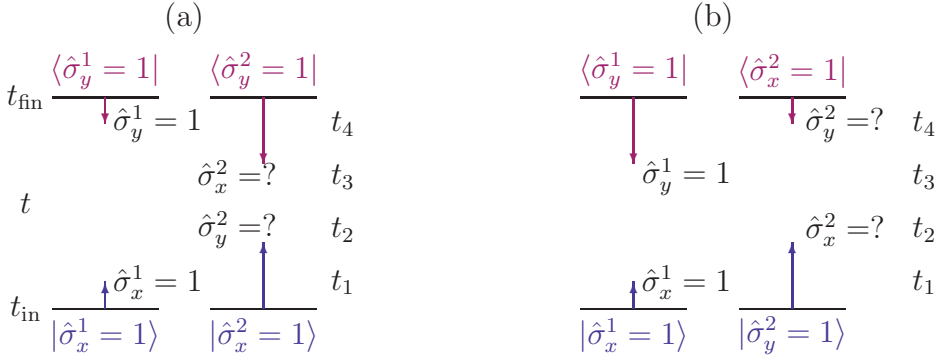
To see this, consider that $\hat{\sigma}^1_x\hat{\sigma}^2_y\hat{\sigma}^2_x\hat{\sigma}^1_y$ corresponds to the sequence of measurements
represented in figure \ref{2vmermin2}.a.  While the pre-selection of
particle 2 is $\hat{\sigma}^2_x=1$ at $t_{\mathrm{in}}$, the next measurement after the pre-selection at $t_2$ is for $\hat{\sigma}^2_y$ and only
{\it after} that a measurement of $\hat{\sigma}^2_x$ is performed at $t_3$.  Thus, there is no guarantee  that the $\hat{\sigma}^2_x$  measurement at $t_3$ will give the same  value as the pre-selected state of
$\hat{\sigma}^2_x=1$ or that the $\hat{\sigma}^2_y$ measurement will give the same value as the post-selected state of
$\hat{\sigma}^2_y=1$.
In TSQM, this is  due to the disturbance of the 2-vector boundary conditions which is created by the IM: the initial pre-selected vector $\hat{\sigma}^2_x=1$ from $t_{\mathrm{in}}$ is ``destroyed" when the $\hat{\sigma}^2_y$  measurement at time $t_2$ is performed and therefore cannot inform the later $\hat{\sigma}^2_x$ measurement at time $t_3$. In other words, with the particular PPS  given in fig. \ref{2vmermin}.a and \ref{2vmermin2}.a, the operator, 
$\hat{\sigma}^1_x\hat{\sigma}^2_y\hat{\sigma}^2_x\hat{\sigma}^1_y$ depends on information from both the
pre-selected vector $\hat{\sigma}^1_x=1$, $\hat{\sigma}^2_x=1$ and the post-selected vector
$\hat{\sigma}^1_y=1$, $\hat{\sigma}^2_y=1$ in a diagonal-PPS sense.   We call this diagonal-PPS because a line connecting $\hat{\sigma}^1_x(t_1)$ with $\hat{\sigma}^2_x(t_3)$ will be diagonal or will cross the line connecting $\hat{\sigma}^2_y(t_2)$ with $\hat{\sigma}^1_y(t_4)$, where $t_{\mathrm{in}}<t_1<t_2 ...<t_{\mathrm{fin}}$, see fig. \ref{2vmermindiag}.a).   

These results can also be seen in an actual measurement situation, we consider interaction Hamiltonians with coupling terms  $\hat{\sigma}^1_x\delta (t-t_1)$, $\hat{\sigma}^2_x\delta (t-t_2)$, $\hat{\sigma}^2_y\delta (t-t_2)$ and $\hat{\sigma}^1_y\delta (t-t_1)$: 
\begin{eqnarray}
  \label{psi22}
&& \langle\hat{\sigma}^1_y=1\vert\langle \hat{\sigma}^2_y=1\vert e^{i\hat{Q}_1\hat{\sigma}^1_x\hat{\sigma}^2_x}e^{i\hat{Q}_2\hat{\sigma}^1_y\hat{\sigma}^2_y}
\vert\hat{\sigma}^1_x=1\rangle\vert\hat{\sigma}^2_x=1\rangle\nonumber\\
&=& \langle\hat{\sigma}^1_y=1\vert\langle \hat{\sigma}^2_y=1\vert e^{i\hat{Q}_2\hat{\sigma}^1_y\hat{\sigma}^2_y}e^{i\hat{Q}_1\hat{\sigma}^1_x\hat{\sigma}^2_x}
\vert\hat{\sigma}^1_x=1\rangle\vert\hat{\sigma}^2_x=1\rangle
=1
\end{eqnarray}
Since $e^{i\hat{Q}_1\hat{\sigma}^1_x\hat{\sigma}^2_x}$ commutes with $e^{i\hat{Q}_2\hat{\sigma}^1_y\hat{\sigma}^2_y}$, they can be interchanged and thus the same outcome of $+1$ is obtained.  
However, for  the other observables $\langle\hat{\sigma}^1_y=1\vert\langle \hat{\sigma}^2_y=1\vert e^{i\hat{Q}_1\hat{\sigma}^1_x\hat{\sigma}^2_y}e^{i\hat{Q}_2\hat{\sigma}^1_y\hat{\sigma}^2_x}
\vert\hat{\sigma}^1_x=1\rangle\vert\hat{\sigma}^2_x=1\rangle$
the opposite eigenvalue is obtained (even though they commute)  i.e. 
$\hat{\sigma}^1_x\hat{\sigma}^2_y\hat{\sigma}^2_x\hat{\sigma}^1_y$ will give an outcome of $-1$ given the PPS of fig. \ref{2vmermin2}.a even though separately $\hat{\sigma}^1_x\hat{\sigma}^2_y=+1$ and $\hat{\sigma}^2_x\hat{\sigma}^1_y=+1$.  This is thus a violation of the product rule (see fig. \ref{2vmermindiag}.a).  
This diagonal-PPS phenomenon can be generalized to functions that are polynomials of products of observables with the proper ordering (i.e. no mixing or sandwiching).

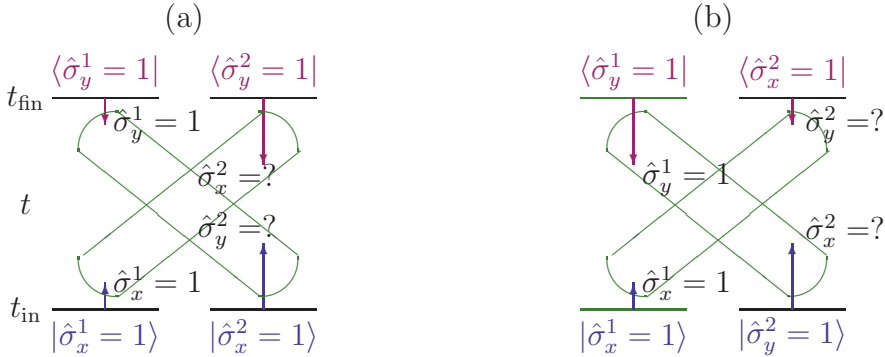
\begin{figure}[h] \epsfxsize=4.5truein
\vskip 1.2cm
\begin{picture}(400,90)(0,0)
\put(40,10){\line(1,0){40}}
\put(40,90){\line(1,0){40}}
\color{BlueViolet}
\put(60,10){\vector(0,1){10}}
\color{RedViolet}
\put(60,90){\vector(0,-1){10}}

\color{Black}
\put(100,10){\line(1,0){40}}
\put(100,90){\line(1,0){40}}
\color{BlueViolet}

\put(120,10){\vector(0,1){25}}
\color{RedViolet}

\put(120,90){\vector(0,-1){25}}
\color{OliveGreen}

\color{Black}
\color{OliveGreen}

\put(65,70){\oval(30,30)[tl]}
\put(50,70){\line(5,-4){69}}
\put(118,30){\oval(30,30)[br]}
\put(65,85){\line(5,-4){68}}

\put(118,70){\oval(30,30)[tr]}
\put(134,70){\line(-5,-4){69}}
\put(65,30){\oval(30,30)[bl]}
\put(120,85){\line(-5,-4){68}}

\put(265,70){\oval(30,30)[tl]}
\put(250,70){\line(5,-4){69}}
\put(318,30){\oval(30,30)[br]}
\put(265,85){\line(5,-4){68}}

\put(318,70){\oval(30,30)[tr]}
\put(334,70){\line(-5,-4){69}}
\put(265,30){\oval(30,30)[bl]}
\put(320,85){\line(-5,-4){68}}

\put(240,10){\line(1,0){40}}
\put(240,90){\line(1,0){40}}
\color{BlueViolet}

\put(260,10){\vector(0,1){10}}
\color{RedViolet}

\put(260,90){\vector(0,-1){25}}
\color{OliveGreen}

\color{BlueViolet}
\put(260,0){\makebox(0,0){$|\hat{\sigma}_x^1=1\ra$}}
\put(120,0){\makebox(0,0){$|\hat{\sigma}_x^2=1\ra$}}
\put(60,0){\makebox(0,0){$|\hat{\sigma}_x^1=1\ra$}}
\put(320,0){\makebox(0,0){$|\hat{\sigma}_y^2=1\ra$}}

\color{RedViolet}
\put(260,100){\makebox(0,0){$\la\hat{\sigma}_y^1=1|$}}
\put(120,100){\makebox(0,0){$\la\hat{\sigma}_y^2=1|$}}
\put(60,100){\makebox(0,0){$\la\hat{\sigma}_y^1=1|$}}
\put(320,100){\makebox(0,0){$\la\hat{\sigma}_x^2=1|$}}
\color{Black}
\put(300,10){\line(1,0){40}}
\put(300,90){\line(1,0){40}}
\color{BlueViolet}

\put(320,10){\vector(0,1){25}}
\color{RedViolet}

\put(320,90){\vector(0,-1){10}}
\color{Black}
\put(30,10){\makebox(0,0){$t_{\mathrm{in}}$}}
\put(30,90){\makebox(0,0){$t_{\mathrm{fin}}$}}
\put(30,50){\makebox(0,0){$t$}}
\put(90,120){\makebox(0,0){(a)}}
\put(290,120){\makebox(0,0){(b)}}

\put(80,20){\makebox(0,0){$\hat{\sigma}^1_x=1$}}
\put(110,40){\makebox(0,0){$\hat{\sigma}^2_y=?$}}
\put(110,60){\makebox(0,0){$\hat{\sigma}^2_x=?$}} 
\put(80,80){\makebox(0,0){$\hat{\sigma}^1_y=1$}}

\put(280,20){\makebox(0,0){$\hat{\sigma}^1_x=1$}}
\put(340,40){\makebox(0,0){$\hat{\sigma}^2_x=?$}}
\put(280,60){\makebox(0,0){$\hat{\sigma}^1_y=1$}} 
\put(340,80){\makebox(0,0){$\hat{\sigma}^2_y=?$}}

\end{picture}

\caption{\small a) Measurement of $\hat{\sigma}^1_x\hat{\sigma}^2_y\hat{\sigma}^2_x\hat{\sigma}^1_y$ is diagonal, b) measurement of $\hat{\sigma}^1_x\hat{\sigma}^2_x\hat{\sigma}^1_y\hat{\sigma}^2_y$ is diagonal}
\label{2vmermindiag}
\end{figure}

To summarize this sub-section, given the PPS of fig. \ref{2vmermindiag}.a, the subset of observables circled in fig. 
6.a  (and the products of those circled observables) can be assigned eigenvalues in a way that satisfies the function relation requirement eq. \ref{KS1}. 
But, the product of the other observables (e.g. $\hat{\sigma}^1_x\hat{\sigma}^2_y$ and $\hat{\sigma}^2_x\hat{\sigma}^1_y$) can only be ascertained (given this particular PPS) using information from {\it both} the pre- and post-selected vector in a diagonal sense (see fig. \ref{2vmermindiag}.a), and will thus violate the product rule.  With the PPS  of  fig. \ref{2vmermindiag}.b, the subset in fig. 
6.b (and the relevant products of observables) can be assigned eigenvalues in a way that satisfies the function relation requirement eq. \ref{KS1}.  But, the product of the other observables $\hat{\sigma}^1_x\hat{\sigma}^2_x$ and $\hat{\sigma}^1_y\hat{\sigma}^2_y$, i.e. $\hat{\sigma}^1_x\hat{\sigma}^2_x\hat{\sigma}^1_y\hat{\sigma}^2_y$ violate the function rule eq. \ref{KS1}.

\vskip -1pc

\singlespacing
\begin{figure}[h]
\vskip -2pc
\begin{eqnarray*}
& {\hat{\sigma}_x}^1  \;\;\;\;\;\;\;\;\;\;\;\; {\hat{\sigma}_x}^2 
  \;\;\;\;\;\;\; {\hat{\sigma}_x}^1{\hat{\sigma}_x}^2 \;\; \;\;\;   \;\; \;\;\;  \;\; \;\;\;   \;\; \;\;\; \;\; \;\;\;   \;\; \;\;\; {\hat{\sigma}_x}^1  \;\;\;\;\;\;\;\;\;\;\;\; {\hat{\sigma}_x}^2 
  \;\;\;\;\;\;\; {\hat{\sigma}_x}^1{\hat{\sigma}_x}^2 \;\; \;\;\;\nonumber \\
\nonumber \\
& {\hat{\sigma}_y}^1  \;\;\;\;\;\;\;\;\;\;\;\; {\hat{\sigma}_y}^2 
  \;\;\;\;\;\;\; {\hat{\sigma}_y}^1{\hat{\sigma}_y}^2 \;\; \;\;\; \;\; \;\;\;  \;\; \;\;\;   \;\; \;\;\; \;\; \;\;\;   \;\; \;\;\; {\hat{\sigma}_y}^1  \;\;\;\;\;\;\;\;\;\;\;\; {\hat{\sigma}_y}^2 
  \;\;\;\;\;\;\; {\hat{\sigma}_y}^1{\hat{\sigma}_y}^2 \;\; \;\;\;\nonumber \\ 
\nonumber \\
& {\hat{\sigma}_x}^1{\hat{\sigma}_y}^2  \;\;\;\;\;  \;\;  {\hat{\sigma}_x}^2 {\hat{\sigma}_y}^1
 \;\; \;\;\; {\hat{\sigma}_z}^1{\hat{\sigma}_z}^2\;\; \;\;\;   \;\; \;\;\;  \;\; \;\;\;   \;\; \;\;\; \;\; \;\;\;   \;\; \;\;\;{\hat{\sigma}_x}^1{\hat{\sigma}_y}^2  \;\;\;\;\;   \;\; {\hat{\sigma}_x}^2 {\hat{\sigma}_y}^1
 \;\; \;\;\; {\hat{\sigma}_z}^1{\hat{\sigma}_z}^2\;\; \;\;\; \nonumber \\ 
\nonumber \\
&\nonumber\\
\end{eqnarray*}
\vskip -1pc

\begin{picture}(400,90)(0,0)
\put(90,120){\makebox(0,0){(a)}}
\put(320,120){\makebox(0,0){(b)}}

\color{BlueViolet}

\put(105,187){\oval(120,20)}
\put(105,222){\oval(120,20)}

\put(155,180){\oval(39,104)}

\put(325,183){\oval(39,104)}
\put(275,183){\oval(39,104)}

\put(319,149){\oval(120,20)}

\color{Red}

\label{2merm4dov14}

\end{picture}
\vskip -8pc

\vskip -2pc
\caption[]{Products of observables that are not disturbed, a) given the PPS of fig. 2.a, and b) given the PPS of fig. 2.b}
\end{figure}
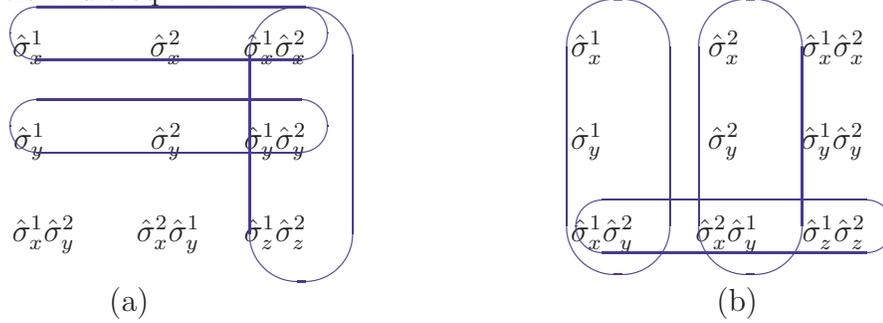
\vskip -2pc

\subsubsection{\bf Ascertaining the results of any one of 16 observables through the generalized state: }
\label{ascgs}

As was explained in \S \ref{asc9}, definite results for any {\it one} of the intermediate measurements can be obtained for several different PPSs which were complete measurements (and thus describable by a wavefunction).  In addition, some triplets (products of observables) can also be ascertained (see fig. 6) given a particular pre- or post-selection.  However, for the most general setup considered in this section, no 2 products of observables can be ascertained.  The general setup considers superpositions of these PPSs.  In fact, this is required in order  to ascertain any one of the $16$ observables in Mermin's successful generalization of VAA (it is also required to ascertain any one of the $9$ for general PPSs). 
Following VAA  \cite{s-xyz}, Mermin showed that the way that any one of the $16$ values  of the 7 BKS nonets (e.g. fig. 2 is one of them, the others given by $\sigma_\nu\sigma_\mu$)

can be ascertained with certainty is by entangling the 2-particle system representing the 4D BKS nonet (represented by the pre and post-selection of fig. \ref{2vmermin}.a and fig. \ref{2vmermin}.b, etc., and labeled as $|i\rangle$)  with another system, i.e. an ancilla (represented here by $|\Psi_i\rangle$ and $|\Phi_i\rangle$).  
The nonet  is prepared at $t_{\mathrm{in}}$ by correlating it with a 
set of states of an ancilla:
\begin{equation}
 |A\rangle=\sum_i \frac{1}{\sqrt{N}}|\Phi_i\rangle|i\rangle
\end{equation}
Then the ancilla is ``guarded" so there are no interactions with the ancilla during the time
$(t_{\mathrm{in}}, t_{\mathrm{fin}})$. At $t_{\mathrm{fin}}$ we post select  on the particle and ancilla and obtain the state:
\begin{equation}
|\Omega\rangle=\frac{1}{\sqrt{N}}\sum_i|\Psi_i\rangle|i\rangle
\end{equation}
  If we
are successful in obtaining this state for the post-selection, then the state of the system is
described in the intermediate
time  by the entangled state (see figure \ref{mult2particles})~\cite{jmav, ar}

 \begin{equation}
\Psi = \sum _{i} \alpha _{i} \langle \Psi _{i} \mid \mid \Phi _{i}
\rangle
\label{genstate}
\end{equation}

\vskip -1.5cm
\begin{figure}[h] \epsfxsize=3.5truein

\begin{picture}(400,130)(0,0)
\put(40,10){\line(1,0){40}}
\put(40,90){\line(1,0){40}}
\color{BlueViolet}
\put(60,10){\vector(0,1){20}}
\color{RedViolet}
\put(60,90){\vector(0,-1){20}}
\color{OliveGreen}

\put(60,50){\oval(25,80)}

\put(55,37){\makebox(0,0){\bf \small correlated}}
\put(95,50){\makebox(0,0){\bf \huge +}}

\color{Black}
\put(100,10){\line(1,0){40}}
\put(100,90){\line(1,0){40}}
\color{BlueViolet}

\put(120,10){\vector(0,1){20}}
\color{RedViolet}

\put(120,90){\vector(0,-1){20}}
\color{OliveGreen}

\put(320,50){\oval(25,80)}

\put(120,37){\makebox(0,0){\bf \small correlated}}
\put(155,50){\makebox(0,0){\bf \huge +}}
\put(300,50){\makebox(0,0){\bf \huge +}}

\color{Black}
\put(160,10){\line(1,0){40}}
\put(160,90){\line(1,0){40}}
\color{BlueViolet}

\put(180,10){\vector(0,1){20}}
\color{RedViolet}

\put(180,90){\vector(0,-1){20}}
\color{OliveGreen}
\put(120,50){\oval(25,80)}
\put(180,50){\oval(25,80)}

\color{BlueViolet}
\put(180,0){\makebox(0,0){$|\Phi_3\ra$}}
\put(120,0){\makebox(0,0){$|\Phi_2\ra$}}
\put(60,0){\makebox(0,0){$|\Phi_1\ra$}}
\put(320,0){\makebox(0,0){$|\Phi_n\ra$}}

\color{RedViolet}
\put(180,100){\makebox(0,0){$\la\Psi_3|$}}
\put(120,100){\makebox(0,0){$\la\Psi_2|$}}
\put(60,100){\makebox(0,0){$\la\Psi_1|$}}
\put(320,100){\makebox(0,0){$\la\Psi_n|$}}
\color{Black}
\put(210,50){\circle*{3}}
\put(230,50){\circle*{3}}
\put(250,50){\circle*{3}}
\put(270,50){\circle*{3}}
\put(300,10){\line(1,0){40}}
\put(300,90){\line(1,0){40}}
\color{BlueViolet}

\put(320,10){\vector(0,1){20}}
\color{RedViolet}

\put(320,90){\vector(0,-1){20}}
\color{Black}
\put(30,10){\makebox(0,0){$t_{\mathrm{in}}$}}
\put(30,90){\makebox(0,0){$t_{\mathrm{fin}}$}}
\put(30,50){\makebox(0,0){$t$}}
\end{picture}

\caption[Generalized State: superpositions of 2-vectors]
{Generalized State: superpositions of 2-vectors given by eq. \ref{genstate}.}
\label{mult2particles}
\end{figure}
For general PPSs, we use multiple sets of boundary conditions given by fig. \ref{2vmermin}.a and fig. \ref{2vmermin}.b, etc.,
to get an entangled state represented by fig. \ref{2vmermin22} (where for simplicity we have taken the states of the ancilla to be an orthonormal set, $\mathrm{\bf f}^\mu$).  Mermin then presented an elegant method to determine the states of the ancilla necessary to produce the effect: he selected a definite representation for the 2-particle spins, performed a projection, $\mathrm{\bf \hat{P}}$, onto the subspace given by the nonet, and solved under rotation for the state of the ancilla.  That is, 
$\langle\Omega|\mathrm{\bf \hat{P}}|A\rangle=0$
 for ``all but a single one of the projections associated with..."~\cite{mermin2v} the observable of the nonet that is to be ascertained with certainty.  

\begin{figure}[h] \epsfxsize=4.5truein
\begin{picture}(400,60)(0,0)
\put(40,10){\line(1,0){40}}
\put(40,50){\line(1,0){40}}
\color{BlueViolet}
\put(60,10){\vector(0,1){15}}
\color{RedViolet}
\put(60,50){\vector(0,-1){15}}

\color{Black}
\put(100,10){\line(1,0){40}}
\put(100,50){\line(1,0){40}}
\color{BlueViolet}

\put(120,10){\vector(0,1){15}}
\color{RedViolet}

\put(120,50){\vector(0,-1){15}}
\color{OliveGreen}

\color{Black}
\put(240,10){\line(1,0){40}}
\put(240,50){\line(1,0){40}}
\color{BlueViolet}

\put(260,10){\vector(0,1){15}}
\color{RedViolet}

\put(260,50){\vector(0,-1){15}}
\color{OliveGreen}
\color{BlueViolet}
\put(260,0){\makebox(0,0){$|\hat{\sigma}_x^1=1\ra$}}
\put(120,0){\makebox(0,0){$|\hat{\sigma}_x^2=1\ra$}}
\put(60,0){\makebox(0,0){$|\hat{\sigma}_x^1=1\ra$}}
\put(320,0){\makebox(0,0){$|\hat{\sigma}_y^2=1\ra$}}

\color{RedViolet}
\put(260,60){\makebox(0,0){$\la\hat{\sigma}_y^1=1|$}}
\put(120,60){\makebox(0,0){$\la\hat{\sigma}_y^2=1|$}}
\put(60,60){\makebox(0,0){$\la\hat{\sigma}_y^1=1|$}}
\put(320,60){\makebox(0,0){$\la\hat{\sigma}_x^2=1|$}}
\color{Black}
\put(300,10){\line(1,0){40}}
\put(300,50){\line(1,0){40}}
\color{BlueViolet}

\put(320,10){\vector(0,1){15}}
\color{RedViolet}

\put(320,50){\vector(0,-1){15}}
\color{Black}
\put(0,10){\makebox(0,0){$t_{\mathrm{in}}$}}
\put(0,50){\makebox(0,0){$t_{\mathrm{fin}}$}}
\put(0,30){\makebox(0,0){$t$}}

\put(170,5){\makebox(0,0){ $\bf \bigotimes \!\!\mid\! { \mathrm{\bf f}^0}\rangle$}}

\put(170,55){\makebox(0,0){ $\bf \bigotimes \!\!\mid\! { \mathrm{\bf f}^0}\rangle$}}

\put(55,5){\makebox(0,0){{\Huge $\,\,\,\,\,\,\,\,\,\,\,\,\,\,\,\,\,\,\,\, \left\{ \,\,\,\,\,\,\,\,\,\,\,\,\,\,\,\,\,\,\,\,\,\,\,\, \right\}$}}}

\put(55,55){\makebox(0,0){{\Huge $\,\,\,\,\,\,\,\,\,\,\,\,\,\,\,\,\,\,\,\, \left\{ \,\,\,\,\,\,\,\,\,\,\,\,\,\,\,\,\,\,\,\,\,\,\,\, \right\}$}}}

\put(200,5){\makebox(0,0){\bf \Huge +}}

\put(200,55){\makebox(0,0){\bf \Huge +}}

\put(375,5){\makebox(0,0){ $\bf \bigotimes \!\!\mid\! { \mathrm{\bf f}^1}\rangle$}}
\put(375,55){\makebox(0,0){ $\bf \bigotimes \!\!\mid\! { \mathrm{\bf f}^1}\rangle$}}

\put(260,5){\makebox(0,0){{\Huge $\,\,\,\,\,\,\,\,\,\,\,\,\,\,\,\,\, \left\{ \,\,\,\,\,\,\,\,\,\,\,\,\,\,\,\,\,\,\,\,\,\,\,\,\,\,\, \right\}$}}}

\put(260,55){\makebox(0,0){{\Huge $\,\,\,\,\,\,\,\,\,\,\,\,\,\,\,\,\, \left\{ \,\,\,\,\,\,\,\,\,\,\,\,\,\,\,\,\,\,\,\,\,\,\,\,\,\,\, \right\}$}}}

\put(410,5){\makebox(0,0){\bf \Huge +...}}

\put(410,55){\makebox(0,0){\bf \Huge +...}}

\end{picture}

\caption{\small Generalized state for BKS nonets}
\label{2vmermin22}
\end{figure}
As Mermin proved, all 4 components of this generalized state are necessary (i.e. we only determine a Bell-state at $t_{\mathrm{fin}}$ on the ancilla rather than make a projection onto any given component $\mathrm{\bf f}^\mu$) to  ascertain a definite answer to any one of the individual observables.

\subsubsection{\bf A physical reason for restrictions on these assignments}
\label{dist}
We have suggested  a physical reason based on TSQM and PPS for the 2 different
values for $\hat{\sigma}^1_z\hat{\sigma}^2_z$.
This points to a  physical reason why no 2 measurements can be ascertained with certainty in the intermediate\footnote{If we were considering a single PPS, as discussed in \S \ref{asc9} and \S\ref{ascprod}, then {\it some} pairs of products of observables can be ascertained with certainty, but not any 2 pairs.
  No 2 products can be ascertained in the case of 16 observables and some 2 pairs of products cannot be ascertained for the 9 observables.} time:
all sets of boundary conditions are needed (those corresponding to both \ref{2vmermin}.a and  \ref{2vmermin}.b, etc.) in order to ascertain with certainty the value of any one of the 16 observables, as represented by fig. \ref{2vmermin22}.  However, when the first observable is ascertained,
then it will depend on both the pre- and the post-selection measurement (i.e. it will be diagonal-PPS) in 2 of the 4 components of the generalized state (see \S \ref{ascprod}) and  
 will collapse the entire configuration onto a subset of the PPSs, thereby  disturbing the terms of the generalized state. Given any pair of measurements, there will always be a diagonal-PPS situation when all 4 components of the generalized state are considered.
This can be seen by  comparing figs. 6.a and 6.b and noting that any 2 observables will not be circled in both. Therefore, since we cannot be sure that the entire setup (see fig. \ref{2vmermin22}) is not disturbed, we cannot ascertain with certainty the outcome for any one of the 16 observables for the second measurement.
Furthermore, this arrangement is the maximal correlation that can be performed (i.e. the 4D state of 2 spins can be maximally correlated to another 4D system as performed here), and thus we cannot create an even more sophisticated situation with additional ancillas.
We have thus given a physical picture for Mermin's ``intriguing" question: there will always be a diagonal situation for any 2 observables.  

\bigskip

\subsection{\bf Non-classical Weak Values for the 4D BKS nonets}
\label{bkswm}
We can now clarify Mermin's statement: ``Alice's other two `results' have nothing to do with any properties of the particle or the results of any measurement actually performed."~\cite{mermin2v}  While it is certainly true that these ``other results" cannot be ascertained simultaneously in terms of an IM (as was demonstrated in \S \ref{dist} and by Mermin), they {\bf can} be measured simultaneously through WMs. 

The route to an easy calculation of WVs can be established from 
Mermin's description of VAA's accomplishment: ``Alice's list gives the observed result for the measurement Bob actually made and had he measured anything else it would have given the result he observed."~\cite{mermin2v} This provides a direct route to WMs through theorem 2: WMs will produce the identical result as predicted for the IM since the IM results are definite.  Thus,  the other `results' are related to properties of the particle and can be simultaneously measured.
\label{wvnonet}

We can also obtain non-classical results in this example (similar to the 3-box-paradox), by first re-writing the observables of fig. 2 in terms of 3 spin components for 2 ``virtual" particles:  for the first particle, 
$\mathrm{\bf \hat{S}^1_3}\equiv \hat{\sigma}_x^1\hat{\sigma}_y^2$, 
$\mathrm{\bf \hat{S}^1_2}\equiv \hat{\sigma}_x^1 \hat{\sigma}_z^2$, and 
$\mathrm{\bf \hat{S}^1_1}\equiv \hat{\sigma}_x^2$; for particle 2, components which commute with particle 1 are 
$\mathrm{\bf \hat{S}^2_3}\equiv \hat{\sigma}_x^2{\hat{\sigma}_y}^1$, 
$\mathrm{\bf \hat{S}^2_2}\equiv \hat{\sigma}_z^1 {\hat{\sigma}_x}^2$, and 
$\mathrm{\bf \hat{S}^2_1}\equiv \hat{\sigma}_x^1$.  
We can observe a non-classical WV by noting that 
$\hat{\sigma}^1_x\hat{\sigma}^2_y\hat{\sigma}^2_x\hat{\sigma}^1_y=-1$  given the PPS of fig. \ref{2vmermin2}.a even though separately $\hat{\sigma}^1_x\hat{\sigma}^2_y=+1$ and $\hat{\sigma}^2_x\hat{\sigma}^1_y=+1$, i.e. a violation of the product rule and thus a diagonal situation (see fig. \ref{2vmermindiag}.a).  Thus WMs 
must yield the same outcomes, i.e. $\mathrm{\bf \hat{N}}(\mathrm{\bf \hat{S}^1_3})_{\mathrm{w}}=+1$ and 
$\mathrm{\bf \hat{N}}(\mathrm{\bf \hat{S}^2_3})_{\mathrm{w}}=+1$ but $\mathrm{\bf \hat{N}}(\mathrm{\bf \hat{S}^1_3}\mathrm{\bf \hat{S}^2_3})_{\mathrm{w}}=-1$, a non-classical result.  To analyze these results, we define the following pair occupation operators:

\bigskip
\noindent $\mathrm{\bf \hat{N}}_{++}$ the projector on the state $\mathrm{\bf \hat{S}^1_3}=1$, and 
$\mathrm{\bf \hat{S}^2_3}=1$

\noindent $\mathrm{\bf \hat{N}}_{+-}$ the projector on the state $\mathrm{\bf \hat{S}^1_3}=1$, and 
$\mathrm{\bf \hat{S}^2_3}=-1$

\noindent $\mathrm{\bf \hat{N}}_{-+}$ the projector on the state $\mathrm{\bf \hat{S}^1_3}=-1$, and 
$\mathrm{\bf \hat{S}^2_3}=1$

\noindent $\mathrm{\bf \hat{N}}_{--}$ the projector on the state $\mathrm{\bf \hat{S}^1_3}=-1$, and 
$\mathrm{\bf \hat{S}^2_3}=-1$. 
\bigskip

\noindent We can relate these measurements to the 3 box example of \S \ref{3boxesintro}, but in this case we have 2 boxes and 2 particles:  $\mathrm{\bf \hat{N}}_{++}$ means the number of times that particle 1 and particle 2 are in the first box, $\mathrm{\bf \hat{N}}_{+-}$ means the number of times that particle 1 is in the first box and particle 2 is in the second box, etc. 

The WV of the projection operator $(1-\mathrm{\bf \hat{S}^1_3})(1-\mathrm{\bf \hat{S}^2_3})$ is $-\frac{1}{2}$, a non-classical result.
From Theorems 1 and 2, 
we can deduce the following: 
the different ways to obtain $\mathrm{\bf \hat{N}}(\mathrm{\bf \hat{S}^2_3})=+1$ are given by $\mathrm{\bf \hat{N}}_{++}$ (i.e. $\mathrm{\bf \hat{S}^1_3}=1$, and $\mathrm{\bf \hat{S}^2_3}=1$) and $\mathrm{\bf \hat{N}}_{-+}$ (i.e. $\mathrm{\bf \hat{S}^1_3}=-1$, and 
$\mathrm{\bf \hat{S}^2_3}=1$) and therefore we can deduce that:
\beq
\mathrm{\bf \hat{N}}_{*+}\equiv\mathrm{\bf \hat{N}}_{++}+\mathrm{\bf \hat{N}}_{-+}=1
\label{14dwv1}
\eeq
\noindent In terms of the box analogy, this is how many ways that particle 2 can be found in box 1.   Also $\mathrm{\bf \hat{N}}(\mathrm{\bf \hat{S}^1_3})=+1$, and thus $\mathrm{\bf \hat{N}}(\mathrm{\bf \hat{S}^1_3})\neq -1$ (i.e. how many ways can particle 1 be found in box 1).  This is characterized by:
\beq
\mathrm{\bf \hat{N}}_{+*}\equiv\mathrm{\bf \hat{N}}_{++}+\mathrm{\bf \hat{N}}_{+-}=1
\label{14dwv2}
\eeq
\noindent From $\mathrm{\bf \hat{S}^1_3}\mathrm{\bf \hat{S}^2_3}=-1$ (which again means that both particles cannot be found in the same box), it cannot be that $\mathrm{\bf \hat{S}^2_3}\mathrm{\bf \hat{S}^1_3}=+1$ and thus $\mathrm{\bf \hat{S}^2_3}$ and $\mathrm{\bf \hat{S}^1_3}$ must be opposite in sign.  This can be characterized by:
\beq
\mathrm{\bf \hat{N}}_{++}+\mathrm{\bf \hat{N}}_{--}=0
\label{14dwv3}
\eeq 
Furthermore, since eq. \ref{14dwv1} equals eq. \ref{14dwv2}, we can deduce:
\beq
\mathrm{\bf \hat{N}}_{+-}=\mathrm{\bf \hat{N}}_{-+}
\label{14dwv5}
\eeq
\noindent Subtracting eq. \ref{14dwv3} from  the following identity
\beq
\mathrm{\bf \hat{N}}_{++}+\mathrm{\bf \hat{N}}_{--}+\mathrm{\bf \hat{N}}_{+-}+\mathrm{\bf \hat{N}}_{-+}=+1.
\label{1singwvd}
\eeq
we obtain:
\beq
\mathrm{\bf \hat{N}}_{+-}+\mathrm{\bf \hat{N}}_{-+}=1
\label{14dwv4}
\eeq
\noindent Eq. \ref{14dwv4} implies that the 2 particles are never in the same box.   From eq. \ref{14dwv5} and eq. \ref{14dwv4}, we can deduce:
\beq
\mathrm{\bf \hat{N}}_{-+}=\mathrm{\bf \hat{N}}_{+-}={1\over2}.
\eeq
\noindent Substituting this value into eq. \ref{14dwv2}, we can deduce that:
\beq
\mathrm{\bf \hat{N}}_{++}={1\over2}
\label{1negpair}
\eeq
\noindent Finally, substituting this into eq. \ref{14dwv3}, we can deduce:
\beq
\mathrm{\bf \hat{N}}_{--}=-{1\over2}.
\eeq
As shown in ~\cite{at2}, all these statements can be measured simultaneously through WMs and will yield:
\beq
(\mathrm{\bf \hat{N}}_{-+})_{\mathrm{w}}=(\mathrm{\bf \hat{N}}_{+-})_{\mathrm{w}}=(\mathrm{\bf \hat{N}}_{++})_{\mathrm{w}}={1\over2}.
\label{1pairwv}
\eeq
while:
\beq
(\mathrm{\bf \hat{N}}_{--})_{\mathrm{w}}=-{1\over2}
\label{1negpairwv}
\eeq
In other words, if a WM is performed on the number of times that particle 1 is in the first box and particle 2 is in the first box, then the result is the non-classical result $-\frac{1}{2}$.  
Thus, the way that the 2 seemingly contradictory statements $\hat{\sigma}^1_z\hat{\sigma}^2_z=\pm 1$ can weakly ``peacefully co-exist" (to paraphrase Abner Shimony) is that a WV goes outside the spectrum of possible eigenvalues, i.e. eq. 
\ref{1negpairwv}.  
\bigskip

\noindent In summary, we see again that WMs give an empirical manifestation of BKS:
\begin{itemize}
\item the BKS ``contradiction" here is that $\hat{\sigma}^1_x\hat{\sigma}^2_y\hat{\sigma}^2_x\hat{\sigma}^1_y=-1$  (given the PPS of fig. \ref{2vmermin2}.a) even though separately $\hat{\sigma}^1_x\hat{\sigma}^2_y=+1$ and $\hat{\sigma}^2_x\hat{\sigma}^1_y=+1$
\item these 3 outcomes can be measured weakly without contradiction because the product of WVs is not equal to the WV of the product
\item if BKS were not correct and a noncontextual-HVT were possible, then the product rule should be satisfied and an IM of $\hat{\sigma}^1_x\hat{\sigma}^2_y\hat{\sigma}^2_x\hat{\sigma}^1_y$ should yield $+1$.
This leads to an immediate contradiction because:
\begin{itemize}
\item by theorem 2, the WV must be equal to the ideal result
\item but, this would be inconsistent with an actual WM which will register $(\mathrm{\bf \hat{N}}_{--})_{\mathrm{w}}=-{1\over2}$, 
\end{itemize}
\item therefore, BKS is empirically consistent with WMs 
\end{itemize}
We have thus given a physical explanation for why an IM cannot reveal these values, while WMs can reveal these values. 
Thus, with WMs, the BKS ``contradiction" still exists (i.e. a noncontextual-HVT cannot reproduce QM), yet now it can also be measured.  
In other words, we have physically shown how to obtain
\begin{itemize}
\item $+1$ for the product of all nine observables when this is performed in the sequence of the rows of fig. 2
\item $-1$ for the product of all nine observables when this is performed in the sequence of the columns of fig. 2
\end{itemize}
(assuming that the system is PPS and measured weakly).
The ambiguity in determining whether $\hat{\sigma}^1_z\hat{\sigma}^2_z=+1$ is obtained or $\hat{\sigma}^1_z\hat{\sigma}^2_z=-1$ is obtained gets shifted to the ambiguity of determining which set of boundary conditions is obtained, i.e. it is now a physical property of the system.  

A modification to the experimental setup suggested by ~\cite{szwz} could be used to test the predictions made in this paper.  This setup considers a 4-D Hilbert space represented by two 2-D subsystems, e.g. the path and polarization of a single photon. ~\cite{szwz} find a BKS contradiction for a particular entangled state.  
 If the transverse positions of each photon is used as the pointer, then WMs can be obtained in ~\cite{szwz} with small transverse displacements. The postselected photon distribution then determines the size of these displacements.  Using the technique of ~\cite{at}, WMs can be performed if an optical glass is slightly tilted so that the photon's position is uncertain to within the width of the beam.  
As shown in the next section, the WVs calculated in \S \ref{bkswm} (e.g. eqs. \ref{1negpairwv} and \ref{1pairwv})  are identical to WVs for EPR entanglement~\cite{jt} and thus entanglement in a pre-selected state of 2 particles is isomorphic to an entanglement in our 2 virtual particles.

\subsection{\bf Non-classical weak values for EPR and Peres/BKS}

We shall now show that WVs can show new connections between BKS and EPR.  
Eq. \ref{1negpairwv} and \ref{1pairwv} give the same result as calculating WVs for EPR entanglement~\cite{jt} and thus there is an interesting new kind of isomorphism between the problems of WVs in BKS nonets and EPR.
We can also consider interesting manifestations of the ``contextuality" in these situations by making separate measurements on the ancilla.  Consider a pre-selected  state which is entangled between one of the particles of the nonet and the ancilla (where, following Mermin, we have chosen the ancilla to be an orthonormal set 
$|\mathrm{\bf f}^1\rangle$:
\beq
|A\ra={1\over{\sqrt 2}}(|\hat{\sigma}^1_z=+1\ra|\mathrm{\bf f}^0=+1\ra-|\hat{\sigma}^1_z=-1\ra|\mathrm{\bf f}^0=-1)
\eeq
If we consider product state post-selections, e.g. with $\hat{\sigma}^1_x=1$ and $\mathrm{\bf f}^0=+1$, then we know that an ideal measurement of 
$ \hat{\sigma}^1_x \mathrm{\bf f}^0$ must yield
\beq
\hat{\sigma}^1_x\mathrm{\bf f}^0=1
\label{singwvb}
\eeq
The pre-selected state also yields\footnote{This is easy to see because $\mathrm{\bf f}^1\hat{\sigma}^1_z|\u_z^1\ra|\mathrm{\bf f}^0=-1\ra \rightarrow  |\u_z^1\ra|\mathrm{\bf f}^0=+1\ra$ and $\hat{\sigma}^1_x\mathrm{\bf f}^0|\u_z^1\ra|\mathrm{\bf f}^0=-1\ra \rightarrow  - |\d_z^1\ra|\mathrm{\bf f}^0=-1\ra$ and $\mathrm{\bf f}^1\hat{\sigma}^1_z|\d_z^1\ra|\mathrm{\bf f}^0=+1\ra \rightarrow  - |\d_z^1\ra|\mathrm{\bf f}^0=-1\ra$ and $\hat{\sigma}^1_x\mathrm{\bf f}^0|\d_z^1\ra|\mathrm{\bf f}^0=+1\ra \rightarrow   |\u_z^1\ra|\mathrm{\bf f}^0=+1\ra$.}
\beq
\hat{\sigma}^1_x\mathrm{\bf f}^0+\mathrm{\bf f}^1\hat{\sigma}^1_z=0.
\label{singwva}
\eeq
Thus 
\begin{eqnarray}
&& \lbrace\hat{\sigma}^1_x\mathrm{\bf f}^0+\mathrm{\bf f}^1\sigma^1_z\rbrace \lbrace |\u_z^1\ra|\mathrm{\bf f}^0=-1\ra-|\d_z^1\ra|\mathrm{\bf f}^0=+1\ra\rbrace=\nonumber\\
&& \lbrace - |\d_z^1\ra|\mathrm{\bf f}^0=-1\ra +|\u_z^1\ra|\ra\rbrace -\lbrace |\u_z^1\ra|\ra - |\d_z^1\ra|\mathrm{\bf f}^0=-1\ra\rbrace = 0\nonumber\\
\label{singwvz}
\end{eqnarray}
We can deduce outcomes for un-performed measurements 
of $\hat{\sigma}^1_z$:
\beq
\hat{\sigma}^1_z=-1.
\label{singwvc1}
\eeq
(we have assumed that the ancilla is not disturbed to obtain this conclusion).
We can also deduce outcomes for un-performed measurements 
of $\mathrm{\bf f}^1$ (assuming that the first particle is not disturbed):
\beq
\mathrm{\bf f}^1=-1.
\label{singwvc2}
\eeq
Once again, we see a violation of the product rule~\cite{vaidman1993b}:   from eq. \ref{singwvz} we deduce that $\mathrm{\bf f}^1\hat{\sigma}^1_z=-1$, but if constructed individually from eqs. \ref{singwvc2} and \ref{singwvc1}, $\mathrm{\bf f}^1\hat{\sigma}^1_z=+1$.
This ``conclusion," however, relies on counterfactual statements, 
since not all the required measurements eqs. \ref{singwva}-\ref{singwvc2} can be performed 
simultaneously without disturbing each other. However, we {\bf can} perform WMs on all these statements simultaneously (see \S \ref{bkswm}).

WVs in the EPR situation can also be seen in the instant example if we consider this  PPS (instead of a Bell-state, we  measure a definite state of the ancilla).  We see the identical non-classical results as seen in the previous section.   
First, we define the following projectors

\bigskip
\noindent $\emph{N}_{++}$ the projector on the state $\hat{\sigma}^1_z=1$, and 
$\mathrm{\bf f}^1=1$

\noindent $\emph{N}_{+-}$ the projector on the state $\hat{\sigma}^1_z=1$, and 
$\mathrm{\bf f}^1=-1$

\noindent $\emph{N}_{-+}$ the projector on the state $\hat{\sigma}^1_z=-1$, and 
$\mathrm{\bf f}^1=1$

\noindent $\emph{N}_{--}$ the projector on the state $\hat{\sigma}^1_z=-1$, and 
$\mathrm{\bf f}^1=-1$. 
\bigskip

\noindent  From Theorem 1 and 2, we can deduce the following: from the post-selection $\hat{\sigma}^1_x=+1$, we can deduce that $\mathrm{\bf f}^1=-1$, i.e. eq. \ref{singwvc2}.  The different ways to obtain $\mathrm{\bf f}^1=-1$ are given by $\emph{N}_{+-}$ (i.e. $\hat{\sigma}^1_z=1$, and $\mathrm{\bf f}^1=-1$) and $\emph{N}_{--}$ (i.e. $\hat{\sigma}^1_z=-1$, and 
$\mathrm{\bf f}^1=-1$) and therefore we can deduce the conservation relationship:
\beq
\emph{N}_{+-}+\emph{N}_{--}=1
\eeq
\noindent Now $\hat{\sigma}^1_z=-1$, and thus $\hat{\sigma}^1_z\neq +1$ is characterized by:
\beq
\emph{N}_{+*}\equiv\emph{N}_{++}+\emph{N}_{+-}=0
\label{4dwv2}
\eeq
In terms of the box analogy, this is how many ways that particle 1 can be found in box 1. 
\noindent In addition, if $\mathrm{\bf f}^1=-1$ then $\mathrm{\bf f}^1\neq +1$ thereby giving:
\beq
\emph{N}_{++}+\emph{N}_{-+}=0
\label{4dwv1}
\eeq
That is, in how many ways can particle 2 be found in box 2.
\noindent Furthermore, since eq. \ref{4dwv1} equals eq. \ref{4dwv2}, we can deduce:
\beq
\emph{N}_{+-}=\emph{N}_{-+}
\label{4dwv5}
\eeq
\noindent From the post-selection, we know  that an ideal measurement of $\hat{\sigma}^1_x\mathrm{\bf f}^0$ will yield $+1$, therefore, using eq. \ref{singwva}, we can deduce that $\mathrm{\bf f}^1\hat{\sigma}^1_z=-1$ (which means that both particles cannot be found in the same box), i.e.
\beq
\underbrace{\hat{\sigma}^1_x\mathrm{\bf f}^0}_{=+1}+\underbrace{\mathrm{\bf f}^1\hat{\sigma}^1_z}_{\Rightarrow =-1}=0.
\eeq
\noindent Next we ask how to obtain $\mathrm{\bf f}^1\hat{\sigma}^1_z=-1$.  To obtain this, it cannot be that $\mathrm{\bf f}^1\hat{\sigma}^1_z=+1$ and thus $\mathrm{\bf f}^1$ and $\hat{\sigma}^1_z$ must be opposite in sign.  This can be characterized by:
\beq
\emph{N}_{++}+\emph{N}_{--}=0
\label{4dwv3}
\eeq
\noindent Subtracting eq. \ref{4dwv3} from  the following identity
\beq
\emph{N}_{++}+\emph{N}_{--}+\emph{N}_{+-}+\emph{N}_{-+}=+1.
\label{singwvd}
\eeq
we obtain:
\beq
\emph{N}_{+-}+\emph{N}_{-+}=1
\label{4dwv4}
\eeq
\noindent From eq. \ref{4dwv5} and eq. \ref{4dwv4}, we can deduce:
\beq
\emph{N}_{-+}=\emph{N}_{+-}={1\over2}.
\eeq
\noindent Plugging this value into eq. \ref{4dwv2}, we can deduce that:
\beq
\emph{N}_{++}=-{1\over2}
\label{negpair}
\eeq
\noindent Finally, plugging this into eq. \ref{4dwv3}, we can deduce:
\beq
\emph{N}_{--}={1\over2}.
\eeq
Entanglement in a pre-selected state of 2 particles is isomorphic to an entanglement in our 2 virtual particles.  Thus, if we look at the right variables, then the BKS setup can be seen to be related to EPR.

The situation analyzed above is general and also points to the Peres/BKS-example~\cite{peres}: in summary,  consider a pre-selected  state $|\Psi_{EPR}\ra={1\over{\sqrt 2}}(|\!\u_z^1\ra|\!\d_z^2\ra-|\!\d_z^1\ra|\!\u_z^2\ra)$ for which the following identity holds $\hat{\sigma}^1_x\hat{\sigma}^2_y+\hat{\sigma}^2_x\hat{\sigma}^1_y=0$.  In addition, $|\Psi_{EPR}\ra$ is also an eigenvector (with eigenvalue $-1$) of the following operators $\hat{\sigma}^1_x\hat{\sigma}^2_x$, $\hat{\sigma}^1_y\hat{\sigma}^2_y$, $\hat{\sigma}^1_z\hat{\sigma}^2_z$. It's easy to see that a noncontextual-HVT cannot assign values consistent with these operator relations:
\beq
V_{\vec{\psi}}(\hat{\sigma^1_x}\hat{\sigma^2_x})=V_{\vec{\psi}}(\hat{\sigma^1_y}\hat{\sigma^2_y})
=V_{\vec{\psi}}(\hat{\sigma^1_z}\hat{\sigma^2_z})=-1
\label{eprval}
\eeq
Consider the commuting observables $\hat{A}_1=\hat{\sigma}^1_x\hat{\sigma}^2_y$ and $\hat{A}_2=\hat{\sigma}^2_x\hat{\sigma}^1_y$.  We know that 
\beq
\hat{A}_1\hat{A}_2=\hat{\sigma}^1_x\hat{\sigma}^2_y\hat{\sigma}^2_x\hat{\sigma}^1_y=\hat{\sigma}^2_z\hat{\sigma}^1_z=-1
\label{eprhvt}
\eeq
 in the singlet state. 
The assumption of non-contextuality is that value assignments can be made to eq. \ref{eprhvt}  even when these assignments are taken from a different context: e.g. assigning values from eq. \ref{eprval}, we obtain:
\beq
V_{\vec{\psi}}(\hat{\sigma}^1_x)V_{\vec{\psi}}(\hat{\sigma}^2_y)V_{\vec{\psi}}(\hat{\sigma}^2_x)V_{\vec{\psi}}(\hat{\sigma}^1_y)=(-1)(-1)=+1
\eeq
but experimentally, we obtain $-1$ from eq. \ref{eprhvt}, a contradiction. Thus a noncontextual-HVT is impossible.  However, now we can probe this state by post-selections and obtain $\hat{\sigma}^1_x\hat{\sigma}^2_y=-1$ even though $\hat{\sigma}^1_x=-1$ and $\hat{\sigma}^2_y=-1$.  This, again, can only be done with PPSs since in a pre-selected only system,  for 2 commuting observables, the product rule is satisfied in contrast to PPSs.  In addition, for this type of contextuality with PPSs, either the pre-selected or the post-selected state must be the EPR state (with an additional assumption of invariance under exchange of particles).

\section{\bf PPS AND CONTEXTUALITY IN HIGHER DIMENSIONS}

A future article~\cite{aha} will show how the feat presented in the previous section can be done in higher dimensions,
e.g. to  the GHZ state~\cite{ghzkaf,vaidmanfp}.
In this case we see that it is not possible to  replace the
spin operators by ordinary numbers 
(which is
what a noncontextual-HVT attempts to do).
However, assignments can be correctly made to each of the $\hat{\sigma}$'s if post-selection is utilized.   Once again, the limitations to these assignments can be seen by using the structure of TSQM: if we try to measure all of these observables together, then some of the values will be assigned in  the diagonal-PPS sense, 
and therefore measuring these observables will cause a disturbance even though they commute.

Consider the GHZ case of 3 spins
pre-selected in the state:
\begin{equation}
|\Psi_{in} \rangle = \frac{1}{\sqrt{2}}\ket{\upa_z^1\upa_z^2\upa_z^3}-\frac{1}{\sqrt{2}}\ket{\dwa_z^1\dwa_z^2\dwa_z^3}
\label{ghz}
\end{equation}

Consider that the pre-selected state  is an eigenstate of the following operators:
$\hat{A}_1\equiv \hat{\sigma}_{x}^{1} \hat{\sigma}_{y}^{2} \hat{\sigma}_{y}^{3}$,
$\hat{A}_2\equiv \hat{\sigma}_{y}^{1} \hat{\sigma}_{x}^{2} \hat{\sigma}_{y}^{3}$, 
and $\hat{A}_3\equiv \hat{\sigma}_{y}^{1} \hat{\sigma}_{y}^{2} \hat{\sigma}_{x}^{3}$
with eigenvalue $+1$.  Also, the
  preselected state is an eigenstate of:
\beq
\hat{A}_4\equiv \hat{\sigma}_{x}^{1} \hat{\sigma}_{x}^{2} \hat{\sigma}_{x}^{3}
\label{a4}
\eeq
with eigenvalue, $-1$, and finally 
 $\hat{A}_1 \hat{A}_2 \hat{A}_3=- \hat{A}_4$
However, because $\hat{A}_1$,  $\hat{A}_2$, $\hat{A}_3$, and $\hat{A}_4$ commute, and because $\hat{\sigma}_{x}^{i}$, $\hat{\sigma}_{y}^{j}$, and  $\hat{\sigma}_{y}^{k}$ commute 
one may ask whether it is possible to also satisfy the above equations by replacing the
spin operators by ordinary numbers $\hat{\sigma}^{1}_x = \pm 1 $, $\hat{\sigma}^{1}_y= \pm 1$, which is
what a non-contextual HVT attempts to do. Without post-selection, this cannot be done because the assignments are, once again, inconsistent with the multiplicative structure of the observables because $\hat{A}_1 \hat{A}_2 \hat{A}_3 =1$, a contradiction.

\subsection{\bf GHZ and PPS}
Assignments can be correctly made to each of the $\hat{\sigma}$'s if post-selection is utilized with limitations again arising from diagonal assignments.
E.g., consider a post-selection of the three particles by measuring their $x$ component with $\hat{\sigma}_x = -1 $:
\beq
\langle\Psi_{fin}\vert=\langle\downarrow_x^1 \downarrow_x^2
\downarrow_x^3\vert
\label{postghz}
\eeq

From the post-selected state we {\it strongly} know that $\hat{\sigma}_{x}^{1} \hat{\sigma}_{x}^{2} \hat{\sigma}_{x}^{3}
=-1$ and from the pre-selected state we {\it strongly} know that $\hat{\sigma}_{x}^{i} \hat{\sigma}_{y}^{j} \hat{\sigma}_{y}^{k}=+1$.  Thus statements of the form $\hat{\sigma}_{y}^{j} \hat{\sigma}_{y}^{k}=-1$ can only be made when information is used from both the pre-selected vector {\it and} from the post-selected vector.  For example, if a measurement of $\hat{\sigma}_{y}^{1} \hat{\sigma}_{y}^{2}$ is performed, then we will definitely find $\hat{\sigma}_{y}^{1} \hat{\sigma}_{y}^{2}=-1$.  However, if we attempt to perform a second measurement, e.g. of $\hat{\sigma}_{y}^{3} \hat{\sigma}_{y}^{1}$ then we will not find $\hat{\sigma}_{y}^{3} \hat{\sigma}_{y}^{1}=-1$, because the $\hat{\sigma}_{y}^{1} \hat{\sigma}_{y}^{2}$ measurement will destroy  the pre-selected vector which contains the $\hat{\sigma}_{x}^{i} \hat{\sigma}_{y}^{j} \hat{\sigma}_{y}^{k}=+1$ information  that the $\hat{\sigma}_{y}^{3} \hat{\sigma}_{y}^{1}$ measurement depends on.  This disturbance occurs even though $\hat{\sigma}_{y}^{1} \hat{\sigma}_{y}^{2}$ and $\hat{\sigma}_{y}^{3} \hat{\sigma}_{y}^{1}$ commute!

\label{ghzintro}

\subsection{\bf Weak Values in GHZ state}

We may also consider WMs  of the GHZ observables. 
  With the post-selection $\langle\Psi_{\mathrm{fin}}\vert=\langle\downarrow_x^1 \downarrow_x^2\downarrow_x^3|$, then in the intermediate time we can replace
$\hat{\sigma}_{x}^{1}=\hat{\sigma}_{x}^{2}=\hat{\sigma}_{x}^{3}=-1$ and 
taking the inner product with this post-selection $|\Psi_{\mathrm{fin}}\rangle$, we then find:
\begin{eqnarray}
\label{wmy2y3}
&(\hat{\sigma}^{2}_y \hat{\sigma}^{3}_y)_{\mathrm{w}}\equiv \weakv{\Psi_{\mathrm{fin}}}{\, \hat{\sigma}^{2}_y \hat{\sigma}^{3}_y \, }{\Psi_{\mathrm{in}}}  =  -1 \\
\label{wmy1y2}
&(\hat{\sigma}^{2}_y \hat{\sigma}^{3}_y)_{\mathrm{w}}\equiv \weakv{\Psi_{\mathrm{fin}}}{\, \hat{\sigma}^{1}_y \hat{\sigma}^{2}_y \, }{\Psi_{\mathrm{in}}}  = -1& \\
\label{wmy1y3}
&(\hat{\sigma}^{2}_y \hat{\sigma}^{3}_y)_{\mathrm{w}}\equiv \weakv{\Psi_{\mathrm{fin}}}{\, \hat{\sigma}^{1}_y \hat{\sigma}^{3}_y \, }{\Psi_{\mathrm{in}}}  =  -1&
\end{eqnarray}
Using again the analogy with particles in boxes, eq.(\ref{wmy2y3}) means that particle $2$ and particle $3$ are not together
in the same box, while eq. (\ref{wmy1y2}) means that particle $1$ and particle $2$ are not together
in the same box, and eq. (\ref{wmy1y3}) means that particle $1$ and
particle $3$ are not together in the same box.  But we only have 2
boxes, so if $1$ and $2$ are not in the same box and $1$ and $3$ are not
in the same box, then $2$ and $3$ must be in the same box.  It is clear that the 
above equalities cannot be satisfied  simultaneously by replacing the 
operators for classical numbers taking the values $\pm 1$. 
To simplify this analysis, we define
\beq
(\mathrm{\bf \hat{N}}_{+++})_w = \ket{\upa_y}\bra{\upa_y}_1 \otimes \ket{\upa_y}\bra{\upa_y}_2 \otimes \ket{\upa_y}\bra{\upa_y}_3 \, , 
\eeq
where the 2 
boxes are denoted by $\pm$ referring to the spin component along $y$.  
Using Theorems 1 and 2, it can be shown~\cite{aha} that:
\beq
(\mathrm{\bf \hat{N}}_{+++})_w = \mathrm{\bf \hat{N}}_{---}=-\frac{1}{4}
\label{lll}
\eeq
and 
\beq
(\mathrm{\bf \hat{N}}_{++-})_w = (\mathrm{\bf \hat{N}}_{--+})_w= (\mathrm{\bf \hat{N}}_{+-+})_w = (\mathrm{\bf \hat{N}}_{-+-})_w=...=\frac{1}{4}
\eeq
We have thus shown how to obtain a non-classical negative triplet WVs, which again, cannot be reproduced by a noncontextual-HVT.

\section{\bf CONCLUSION}
Mermin's results reviewed in \S's \ref{asc9}-\ref{dist} were characterized by Mermin as
``...what follows is not idle theorizing about `hidden variables'.  It is a rock solid quantum mechanical effort to answer a perfectly legitimate quantum mechanical question."~\cite{mermin2v}
  BKS showed that noncontextual-HVT's are not possible in general.  
Therefore, an interesting approach to HVT's and BKS is whether anything new can be learned about experimental situations, in the same spirit as Shimony's apt phrase ``experimental metaphysics."
For example, Bell's Theorem led to interesting experiments which tested the notion of whether quantum entanglement could be stronger than classical correlations. 
Another example is Hardy's-paradox~\cite{hardy} (HVTs for position could not be assigned) which was traditionally ``resolved" by arguing that measurements to verify the paradox could not be implemented simultaneously and therefore Hardy's-paradox was purely a formal result without empirical consequences.  However in \cite{at2,jt}, we demonstrated that WMs could  be implemented simultaneously on the paradoxical statements (with experimental results~\cite{stein2}).  E.g. the electron is always on the overlapping path ($N^-_{O w}=1$) and the positron is always on the overlapping path ($N^+_{O w}=1$), yet they are never there together ($N^{+,-}_{O,O w}=0$). Again, non-classical WVs were demonstrated~\cite{at2} by the negative pair-wise occupation, $N^{+,-}_{NO, NO w}=-1$.   This again was a violation of the product rule~\ref{prodrule} ($N^-_{O w}N^+_{O w}\neq N^{+,-}_{O,O w}$) and a manifestation of diagonal-PPSs and non-classical WVs. Nevertheless, despite their non-classical behavior, WVs do obey a simple, intuitive, and, most important, {\it self-consistent} logic. 

In this paper,  we have 
carried this and the program initiated by Mermin further and have 
shown  new ways that the ``charming elementary mathematics" of BKS can manifest empirically.
E.g. in the 3-box-paradox/diagonal-PPS of \S \ref{3boxesintro},
$\mathrm{\bf \hat{P}}_{A}=1$ if only box $A$ is opened, while $\mathrm{\bf \hat{P}}_{\mathrm{B}}=1$ if only box $B$ is opened, but if we measure both box $A$ and box $B$, then the particle will  
not be found in both boxes, i.e. $\mathrm{\bf \hat{P}}_{\mathrm{A}}\mathrm{\bf \hat{P}}_{\mathrm{B}}=0$ even though $\mathrm{\bf \hat{P}}_{\mathrm{A}}$ and $\mathrm{\bf \hat{P}}_{\mathrm{B}}$ commute, a violation of the product rule.  If WMs are performed, then the non-classical result 
$(\mathrm{\bf \hat{P}}_{\mathrm{C}})_{\mathrm{w}}  =-1$ is obtained.
In the Mermin case \S \ref{vaabks}, the violation of the product rule and diagonal-PPS is that $\hat{\sigma}^1_x\hat{\sigma}^2_y\hat{\sigma}^2_x\hat{\sigma}^1_y=-1$  (given the PPS of fig. \ref{2vmermin2}.a) even though separately $\hat{\sigma}^1_x\hat{\sigma}^2_y=+1$ and $\hat{\sigma}^2_x\hat{\sigma}^1_y=+1$ and non-classical WVs are again obtained, $(\mathrm{\bf \hat{N}}_{--})_{\mathrm{w}}=-{1\over2}$.  Similar results were obtained in the 4-D/EPR case (i.e. $\hat{\sigma}^1_x\hat{\sigma}^2_y=-1$ even though $\hat{\sigma}^1_x=-1$ and $\hat{\sigma}^2_y=-1$) and also in the 8-D GHZ case.

How general are these considerations?
Consider any set of commuting observables $\hat{A}_1,\,\hat{A}_2,\,\hat{A}_3$ and any other set of commuting observables $\hat{B}_1,\,\hat{B}_2,\,\hat{B}_3$. Suppose we perform pre-selection measurements on $\hat{A}_1,\,\hat{A}_2,\,\hat{A}_3$ and perform post-selection measurements on $\hat{B}_1,\,\hat{B}_2,\,\hat{B}_3$, and suppose, e.g.  $[\hat{A}_2,\hat{B}_2]\neq 0$.  Products of observables, e.g. $\hat{A}_1\hat{A}_2$, can be ascertained solely based on the pre-selection, while products of observables, e.g.  $\hat{B}_2\hat{B}_3$,  can be ascertained solely based on the post-selection.
However, as soon as we change the sequence of measurements of $\hat{A}_2$ and $\hat{B}_2$, then it is a diagonal-assignment and they will interfere with each other in PPS.
Given this, it is evident that there are versions of BKS 
for which we cannot obtain definite results for IMs utilizing PPSs.  It is not the case that everything that cannot be known by one vector can be known by two vectors: there are many sets of operators for which it is impossible to assign all the possible eigenvalues  via PPS.  
However, our arguments concerning WMs are completely general.

The results of this article could be used to further exploit the differences between classical and quantum information: while the study of the non-classical aspects of entanglement (Einstein-Podolsky-Rosen/Bohm) started as a foundational examination of HVTs, it was subsequently probed experimentally and used as a resource for quantum computation and communication.  
Similarly,  using TSQM  ~\cite{ah2001f}, Englert demonstrated a new form of cryptography \cite{be2001a} and experiments based on the optical version of this problem \cite{en2001} were successfully performed.
With respect to contextuality, BKS is often based solely on the structure of {\it operators}, while quantum information applications  require special {\it states}.  In this article, we have shown how to probe and empirically manifest contextuality through PPS {\it states}, thereby opening new possibilities  of utilizing contextuality in these applications.  

\begin{figure}
\begin{picture}(200,150)(0,0)\epsfxsize=4truein
\put(0,45){\vector(1,0){140}}
\put(50,0){\vector(0,1){140}}
\put(170,40){\makebox(0,0){\small position of particle 2}}
\put(30,150){\makebox(0,0){\small position of }}
\put(30,140){\makebox(0,0){\small particle 1 }}
\put(70,0){\dashbox{1}(0,140)}
\put(0,70){\dashbox{1}(140,0)}
\color{BlueViolet}
\put(65,65){\framebox(10,10)}
\put(120,90){\makebox(0,0){\small particle 1 at $x_1$ and}}
\put(120,75){\makebox(0,0){\small particle 2 at $x_2$}}
\color{RedViolet}
\put(70,45){\circle{10}}
\put(60,45){\makebox(0,0){\small $x_2$}}

\put(50,70){\circle{10}}
\put(50,60){\makebox(0,0){\small $x_1$}}
 \color{Black}
\end{picture}
\caption{\small Weak Value density for 2 particles}
\label{multipart}
\end{figure}
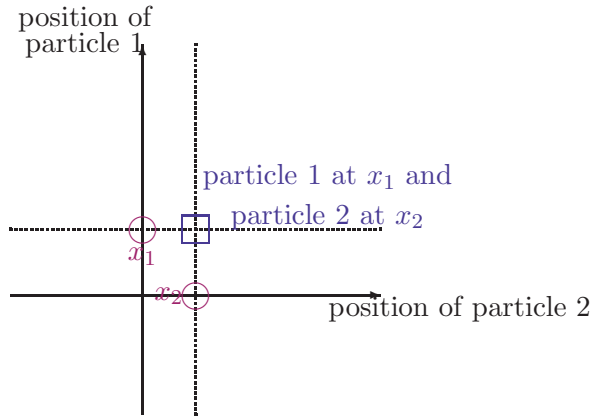

We end with a conjecture: if we start with a non-classical situation as reflected in BKS or a violation of Bell's-inequality, then we can always find a post-selection  which can empirically manifest non-classical WVs.  
  This thereby holds the possibility of shifting foundational debates into a new issue because we can now probe non-classicality in actual WM experiments.   
Non-classical WVs follow intuitively from the logic of WVs: e.g. consider a generic multi-particle systems
in a higher dimensional space $\rho (x_1, x_2)\sim \Psi^*_{fin}(x_1, x_2)\Psi_{in}(x_1, x_2)$ and consider also projections onto space and time.
In general, $\rho (x_1, x_2)$ cannot be measured locally because there is no local way of measuring particle $1$ at $x_1$ simultaneously with measuring particle $2$ at $x_2$:  this would involve a nonlocal Hamiltonian.  However,  the projection on each line can be measured separately:  if it is known that particle $2$ is at $x_2$ then this means $\int \rho (x_1, x_2) dx_1=1$.  Similarly, if it is known that particle $1$ is at $x_1$ then this means $\int \rho (x_1, x_2) dx_2=1$.  However, asking the question if both particles are there together is another point in phase space (see fig. \ref{multipart}).
The particle also could have been located simultaneously in another position with certainty, but then we need to place a negative number somewhere else in order to satisfy the global constraint, i.e. the integral of the WVs has to add up to $1$ (because there is just a single particle along each line), but the individual numbers at each point can be arbitrary.  IM outcomes reveal an integration along just one line, but the WV densities are not just those lines.  E.g. in the Hardy case the seeming  contradiction that both particles are there individually but are not there together
is resolved by a negative number of particles at another point in phase space.  

While there is an operational or experimental meaning to a density over a set of commuting observables, as shown in \S \ref{hvtpps}, 
there is no experimental meaning for a density over a set of non-commuting observables even though such densities may have formal  utility as an aide to calculation~\cite{at6}.
As a result of this dis-connect between densities over non-commuting observables and IMs, there is a similar problem with {\bf WMs and WVs}.  An attempt to give an operational or experimental meaning to WMs of densities over non-commuting observables will illustrate the same problem as demonstrated in \ref{hvtpps}, even though in this article we demonstrated the ability to define WMs and WVs for a {\bf discrete} set of non-commuting observables.

\noindent {\bf Acknowledgments: }
  JT thanks Yakir Aharonov for many conversations and the Templeton foundation for support.

\end{document}